\documentclass[fleqn,usenatbib]{mnras}

\usepackage[T1]{fontenc}
\usepackage{pdflscape}
\DeclareRobustCommand{\VAN}[3]{#2}
\let\VANthebibliography\thebibliography
\def\thebibliography{\DeclareRobustCommand{\VAN}[3]{##3}\VANthebibliography}

\usepackage{graphicx}	
\usepackage{amsmath}	
\usepackage{amssymb}	
\usepackage[export]{adjustbox}

\usepackage{morefloats}
\usepackage{newtxtext,newtxmath}
\usepackage[final]{pdfpages}
\usepackage{subcaption}

\def\h2{H$\rm _2$}

\def\oiii{[O\,{\sc iii}]}

\def\maL{$<t>_L$}
\def\maM{$<t>_M$}
\def\mzL{$<Z>_L$}

\def\kms{$\rm km\,s^{-1}$}
\def\str{{\sc starlight}}
\def\mc{$\mu$m} 
\def\st{{\sc starlight}}

\newcommand\footnoteref[1]{\protected@xdef\@thefnmark{\ref{#1}}\@footnotemark}

\title[The role of recycled gas in AGN feeding]{Observational constraints on the stellar recycled gas in active galactic nuclei feeding}
\author[Riffel, R. et al.]{Rog\'erio Riffel$^{1,2}$\thanks{E-mail: riffel@ufrgs.br},
Luis G. Dahmer-Hahn$^3$, 
Alexandre Vazdekis$^{2,4}$, 
Richard Davies$^5$, 
David Rosario$^6$,
\newauthor
Cristina Ramos Almeida$^{2,4}$,
Anelise Audibert$^{2,4}$,
Ignacio Mart\'\i n-Navarro$^{2,4}$,
Lucimara Pires Martins$^7$,
\newauthor
Alberto Rodr\'\i guez-Ardila$^{8,12,13}$,
Rogemar A. Riffel$^9$,
Thaisa Storchi-Bergmann$^1$,
Michele Bertoldo-Coelho$^1$,
\newauthor
Marina Trevisan$^1$,
Erin Hicks$^{10}$,
Allan Schnorr M\"uller$^1$,
Lais Nery Marinho$^1$ and 
Sylvain Veilleux$^{11}$
\\
$^{1}$ Departamento de Astronomia, Instituto de F\'\i sica, Universidade Federal do Rio Grande do Sul, CP 15051, 91501-970, Porto Alegre, RS, Brazil \\
$^{2}$Instituto de Astrof\'\i sica de Canarias, Calle V\'\i a L\'actea s/n, E-38205 La Laguna, Tenerife, Spain\\
$^3$ Shanghai Astronomical Observatory, Chinese Academy of Sciences, 80 Nandan Road, Shanghai 200030, China\\
$^4$ Departamento de  Astrof\'\i sica, Universidad de La Laguna, E-38205, Tenerife, Spain\\
$^5$ Max-Planck-Institut für extraterrestrische Physik, Postfach 1312, 85741 Garching, Germany \\
$^6$ Centre for Extragalactic Astronomy, Department of Physics, Durham University, South Road, Durham DH1 3LE, UK\\
$^7$ NAT - Universidade Cidade de S\~ao Paulo, Rua Galv\~ao
Bueno, 868, S\~ao Paulo, SP, Brazil. \\
$^8$ Laborat\'{o}rio Nacional de Astrof\'{i}sica - Rua dos Estados Unidos 154, Bairro das Na\c{c}\~{o}es.
CEP 37504-364, Itajub\'{a}, MG, Brazil.\\
$^9$ Universidade Federal de Santa Maria,
Departamento de F\'\i sica/CCNE, 97105-900, Santa Maria, RS, Brazil.\\
$^{10}$  Department of Physics \& Astronomy, University of Alaska Anchorage, Anchorage, AK 99508-4664, USA\\
$^{11}$ Department of Astronomy and Joint Space-Science Institute, University of Maryland, College Park, MD 20742, USA\\
$^{12}$ Observat\'orio Nacional, Rua General Jos\'e Cristino 77, CEP 20921-400, S\~ao Crist\'ov\~ao, Rio de Janeiro, RJ, Brazil\\
$^{13}$ Instituto Nacional de Pesquisas Espaciais, Av. dos Astronautas, 1758 - Jardim da Granja São José dos Campos/SP - 12227-010, Brazil \\
}

\date{Accepted XXX. Received YYY; in original form ZZZ}

\pubyear{}

\begin{document}
\label{firstpage}
\pagerange{\pageref{firstpage}--\pageref{lastpage}}
\maketitle

\begin{abstract}
Near-infrared long-slit spectroscopy has been used to study the stellar population (SP) of the low luminosity active galactic nuclei (AGN) and matched analogues (LLAMA) sample. To perform the SP fits we have employed the X-shooter simple stellar population models together with the \st\ code. Our main conclusions are: The star formation history of the AGNs is very complex, presenting many episodes of star formation during their lifetimes.  In general, AGN hosts have higher fractions of intermediate-age SP (light-weighted mean ages, $<t>_L\lesssim$ 4.5 Gyr) when compared with their analogues ($<t>_L\lesssim$ 8.0 Gyr).  AGN are more affected by reddening and require significant fractions of featureless continuum and hot dust components.  The ratio between the AGN radiated energy and the gravitational potential energy of
the molecular gas ($E_{Rad}$/$E_{PG}$) for the AGN is compared with the \maL\ and a possible anti-correlation is observed. This suggests that the AGN is affecting the star formation in these galaxies, in the sense that more energetic AGN (log$(E_{Rad}$/$E_{PG}) \gtrsim 3$) tend to host nuclear younger SP  ($ <t>_L \lesssim$4Gyr).  We found that the recent ($t<$2~Gyr) returned (recycled) stellar mass is higher in AGN than in the controls.  We also provide evidence that the mass loss of stars would be enough to feed the AGN, thus providing observational constraints for models that predict that AGN feeding is partially due to the recycled gas from dying stars.

\end{abstract}




\section{Introduction}
From the seminal investigations that spotted the correlation between the mass of supermassive black holes (SMBHs) with the velocity dispersion of their host galaxy bulges, it has become accepted that there should be a connection between the active galactic nuclei (AGN) accretion and star formation (SF) processes \citep[e.g.][]{Magorrian+98, Ferrarese+00, Gebhardt+00}. Equally well-established is the coexistence of nuclear SF and AGN activity within the inner regions of galaxies, which strongly suggests a symbiotic relationship between the growth of SMBHs, through gas accretion, and the assembly of galaxies through SF \citep[][for a comprehensively review]{Heckman+14,Madau+14}.

In the galaxy evolution framework, AGN is an important phase in the life cycle of a galaxy, during which its central SMBH accretes material from its vicinity, the circum-nuclear environment (a few hundred pc). The amount of required material to fuel an AGN is small\footnote{To sustain a Seyfert galaxy with a bolometric luminosity of $10^{45} erg s^{-1}$, assuming a characteristic radiative efficiency of accretion of 10 percent, only $\sim$ 0.2\,M$_\odot$/yr are required \citep[see][]{Rosario+18}.} when compared with the amount of material available in the centers of spiral galaxies.  The mechanism responsible for making this material lose angular momentum and ultimately feed the AGN is not a consensus \citep[e.g.][]{Silva-Lima+22}, however, when falling into the deep potential well of an SMBH loses a huge amount of energy. Part of this energy is released in the form of radiation, winds, and relativistic jets. 

Theoretical investigations and numerical simulations examining gas inflows in the vicinity of galactic nuclei have shown that they can give rise to episodes of SF, particularly in the nuclear region \citep[e.g.][]{DiMatteo+05,Hopkins+10,Heckman+14,Zubovas+13,Zubovas+17,Weinberger+23,Mercedes-Feliz+23}. In this context, one of the most frequently invoked processes for regulating SF is AGN feedback \citep[e.g.][ and references therein]{Terrazas+20}. According to some studies, the energy released by the AGN can heat the gas (or remove it) preventing SF \citep[e.g.][]{Granato+04,Zubovas+13,Zubovas+17,Fabian+12,King+15,Trussler+20} or, as suggested in other studies, 
 these outflows and jets, depending on the AGN luminosity, can compress the galactic interstellar medium, and therefore act as a catalyzer, triggering or boosting the SF \citep[e.g.][]{Rees+89,Hopkins+12,Nayakshin+12,Bieri+16,Zubovas+13,Zubovas+17} and even form stars inside the outflow \citep[e.g.][for observational examples, see \citet{Maiolino+17,Gallagher+19}]{Ishibashi+12,Zubovas+13,El-Badry+16,Wang+18}. 
 
 \citet{Mercedes-Feliz+23}, using the cosmological hydrodynamical simulations from the Feedback In Realistic Environment (FIRE), with a novel implementation of hyper-refined accretion-disc winds demonstrate that strong quasar winds, persisting for over 20~Myr, drive the formation of a central gas cavity and significantly reduce the surface density of SF across the galaxy’s disc. The show that the suppression of SF occurs mainly by limiting the availability of gas for SF rather than by ejecting the pre-existing cold gas reservoir (preventive feedback dominates over ejective feedback). These authors also found that locally positive feedback is observed, in the sense that higher local SF efficiency is detected in compressed gas at the edge of the cavity produced by the winds from the central engine. Thus positive and negative feedback are taking place in different galaxy locations \citep[e.g.][]{Bessiere+22}.

From observations, for instance, we have learned that nuclear SF is common in AGN \citep[e.g.][]{Terlevich+90,Storchi-Bergmann+01,CidFernandes+04,Davies+07,Davies+09,Martins+10,Storchi-Bergmann+12,Esquej+14,Ruschel-Dutra+17,Hennig+18,Mallmann+18,Burtscher+21,Dahmer-Hahn+22,Riffel+07,Riffel+09,Riffel+11c,Riffel+16b,Riffel+21,Riffel+22,Riffel+23}. However, comparing the AGN life-cycle  \citep[$\sim$ 0.1 - 10\,Myr,][]{Novak+11,Schawinski+15} with nuclear starbursts ages \citep[$\sim$1-100\,Myr, ][]{Hickox+14} is very difficult, since the uncertainties associated with the latter may be larger than the AGN life-cycle.   Studies have indicated that the proportion of young to intermediate-age stars in the vicinity of AGN is often correlated with the AGN's luminosity. Typically, the most luminous AGN tend to exhibit higher fractions of these younger stellar populations \citep{CidFernandes+04,Riffel+07,Riffel+09,Ellison+16,Bessiere+17,Ruschel-Dutra+17,Zubovas+17,Mallmann+18,Ellison+21,Riffel+21,Riffel+23}. For instance, some studies have found that the age of the starburst is connected to the AGN's luminosity \citep{Davies+07,Riffel+22,Riffel+23}, while no correlation between X-ray luminosity and the fraction of young stellar populations was identified in other analysis \citep{Burtscher+21}. It's worth noting that the complexity of the results found in local AGN studies may partly stem from the luminosity acting as a "third parameter", and the fact that studies on the optical bands may be strongly affected by reddening effects.

The stellar populations, consisting of stars with varying ages, metallicities, and formation histories, are the fundamental building blocks of galaxies. Analyzing these stellar ensembles can offer critical insights into the evolutionary trajectories of galaxies, the dynamics of SF, and the role of feedback mechanisms in shaping the galaxies. Thus it is fundamental to investigate if there is an association between nuclear activity and SF, by studying the stellar population properties in the inner few tens of parsec of AGN over a wide range of AGN luminosities and comparing them with inactive sources with similar host galaxy properties. To disentangle the underlying continuum of AGN is a difficult task since at least 3 main components have to be fitted together, namely: the host galaxy stellar populations (characterized by their ages and metalicities), the AGN accretion disc, and hot dust components \citep[representing the torus emission, for details see][]{Riffel+09}. The wavelength region most sensitive to the AGN featureless components (accretion disc and hot dust emission) and the stellar population at the same time is the near-infrared (NIR$\cong$0.8-2.4\mc) where the AGN featureless continuum (FC), the hot dust (HD) and the stellar population (SP) components can be fitted together and disentangled \citep{Riffel+09,Riffel+11c,Riffel+22}. Additionally, the NIR also offers the possibility to fit the stellar population content in the brighter type~1 sources, which in the optical will have their underlying stellar population out-shined by the accretion disk emission \citep[e.g.][]{Riffel+09}. 

One adequate sample to study the role played by the AGN and star-formation processes is the complete volume-limited sample of nearby AGN (selected by their 14-195 keV luminosity) with its control galaxies called {\it Low Luminosity AGN and Matched Analogues} sample \citep[LLAMA, described in][]{Davies+15}.  The stellar populations in the optical range of the type~2 LLAMA AGN and controls have been studied by \citet{Burtscher+21} who found that the young stellar populations are common inside the nuclear region of the AGN. However, they have shown that these young populations are not indicative of ongoing SF, but rather than to recent cessation in the star-formation providing evidence for models that see AGN activity as a consequence of nuclear SF. Building upon the previous study, here we study the nuclear stellar populations of the entire (type~1, type~2, and controls)  LLAMA sample using NIR spectra, thus offering a more statistically complete analysis in a spectral region less sensitive to dust attenuation and sensitive to the three main components of the AGN spectral energy distribution. This paper is structured as follows: In \S~\ref{sec:data} we present the sample observations and data reduction process. The fitting procedures are described in \S~\ref{popstar}. Results are presented in \S~\ref{sec:results}. We discuss possible implications of the findings in \S~\ref{sec:discussion} and conclusions at \S~\ref{sec:conclusions}.
Along this paper we have adopted in the calculations the galaxies distances listed by \citet{Davies+15} and the \citet{Kroupa+01} initial mass function (IMF).

\section{Sample, Observations and Data Reduction}\label{sec:data}

\subsection{The LLAMA sample of AGN and Control Galaxies}

 The selection of the sample, along with a well-matched control sample, holds the utmost importance in comprehending the impact of AGN activity on SF. Our study focuses on the LLAMA sample \citep{Davies+15}, which consists of 20 AGN meticulously chosen from the {\em Switft}/BAT all-sky ultra-hard X-ray (14-195 keV) survey in its 58-month edition \citep{Ajello+12}. The use of ultra-hard X-rays as a selection criterion presents a significant advantage, as it is mainly unaffected by obscuration, except for the most heavily obscured sources known as Compton-thick \citep[e.g.][]{LaMassa+10,Sales+11,Sales+14,Ricci+15}. According to \citet{Martini+03}, the less luminous the AGN are, the more diverse the fuelling mechanism becomes. To focus on "genuine" AGN, which are more likely to be fuelled by a unified mechanism, we selected the most luminous sources ($\log L_{\rm X} / ({\rm erg/s}) > 42.5$). Moreover, we further refine our selection to include only local AGN ($z < 0.01$), corresponding to distances of approximately $\leq$40~Mpc or less, giving us access to a scale of up to a few hundred parsecs ($\lesssim$200pc).

 Additionally, the LLAMA sample incorporates a control group of 19 galaxies chosen to match the AGN sample in terms of distance, Hubble type, stellar mass (measured by $H$ band luminosity), and axis ratio (inclination). This matched control sample plays a vital role within the LLAMA framework, as it enables us to calibrate our results against a comparable set of galaxies that solely differ from the AGN sample in terms of their nuclear accretion rate. We consider this control sample selection integral to the overall integrity and robustness of the LLAMA project. The main properties of the sample are summarized in  Tab.~\ref{tab:obs_log}. For more details see \citet{Burtscher+21} and \citet{Davies+15}.

\subsection{Observations and data reduction}
Out of our sample, 5 AGN had literature spectra available \citep[taken from][see Tab.~\ref{tab:obs_log} for more details]{Riffel+06,Riffel+19,Mason+15}. A detailed discussion of their observations and reductions can be found in their respective papers, but they followed the standard observation and reduction processes for the NIR (see discussion below for our sample).

The remaining 15 AGN and 19 controls were observed as part of this work, employing either the SpeX spectrograph attached to the Infrared Telescope Facility \citep[IRTF,][]{Rayner+03}, or employing the Triplespec 4 spectrograph, which was mounted on the Victor M. Blanco telescope until 2019 (where it was named ARCoIRIS), or mounted on the SOAR telescope since then (and where it is named TripleSpec 4.1). From now on, we refer to Triplespec 4 as Tspec, regardless of which telescope it was mounted on. During daytime calibrations, we took flat fields and hollow cathode lamps (CuHeAr), in order to correct for discrepancies in sensitivity between pixels and to wavelength calibrate the target. Galaxies were observed using either ABBA pattern (i.e. nodding the object along the slit), or observed in an A-Sky-A pattern, to remove sky emission. Also, each galaxy observed was preceded or followed by an A0V star with similar airmass, which was then used to remove telluric absorptions, as well as to flux calibrate the spectra. Lastly, although we observed hollow arcs during daytime, the objects were $\lambda$-calibrated using sky emission lines, which are more reliable than arcs because they are taken at the same time as the target.

Both SpeX and Tspec standard data reduction is performed using the {\sc spextool} {\sc idl} package  \citep{Vacca+03,Cushing+04}.  This pipeline follows the standard NIR data reduction process and consists of flat-field correction, sky subtraction, a combination of observations to improve the signal-to-noise (S/N) ratio, wavelength calibration, telluric absorptions correction, merging of the different orders, and flux calibration employing the same A0V correction.

\begin{table*}
\centering
\caption{Main properties of our sample and observations.}
\label{tab:obs_log}
\begin{tabular}{lcccccccccr}
\hline
(1) & (2) & (3) & (4) & (5) & (6) & (7) & (8) & (9) & (10) & (11)\\
Source & RA & DEC & K & z & D & Morph. &Instrument/  & Slit Width & Aperture & P.A.  \\
 & (h:m:s) & (d:m:s) & (mag) & & (Mpc) & & Telescope & (arcsec) & (arcsec) & (deg) \\
\noalign{\smallskip}
\hline
\noalign{\smallskip}
\multicolumn{9}{c}{Sample} \\
\noalign{\smallskip}
\hline
\noalign{\smallskip}
NGC 1365      & 03:33:36.4 & -36:08:26.3 & 6.373 & 0.00546 & 18  &Sb    & Tspec/Blanco$^1$ & 1.1 & 2.0 & 0  \\
MCG-05-14-012 & 05:43:32.9 & -27:39:05.0 & 10.80 & 0.00992 & 41  &S0$^+$& Tspec/SOAR$^2$	& 1.1 & 2.0 & 80 \\
NGC 2110      & 05:52:11.3 & -07:27:22.4 & 8.140 & 0.00765 & 34  &S0$^-$& SpeX/IRTF$^5$	& 0.8 & 1.5 & 20  \\
NGC 2992      & 09:45:42.0 & -14:19:34.9 & 8.590 & 0.00771 & 36  &Sa    & SpeX/IRTF$^3$	& 0.8 & 2.0 & 50  \\
MCG-05-23-016 & 09:47:40.1 & -30:56:55.9 & 9.349 & 0.00849 & 35  &S0$^+$& Tspec/SOAR$^2$	& 1.1 & 2.0 & 45  \\
NGC 3081      & 09:59:29.5 & -22:49:34.7 & 8.910 & 0.00798 & 34  &S0/A  & Tspec/Blanco$^4$ & 1.1 & 2.0 & 90  \\
NGC 3783      & 11:39:01.7 & -37:44:19.0 & 8.649 & 0.00973 & 38  &Sab   & Tspec/SOAR$^2$	& 1.1 & 2.0 & 90  \\
NGC 4235      & 12:17:09.8 & +07:11:29.6 & 8.396 & 0.00755 & 37  &Sa    & GNIRS/GEMINI$^6$ & 0.3 & 1.8 & 113  \\
NGC 4388      & 12:25:46.8 & +12:39:43.4 & 8.004 & 0.00842 & 39  &Sb    & GNIRS/GEMINI$^6$ & 0.3 & 1.8 & 64  \\
NGC 4593      & 12:39:39.4 & -05:20:39.0 & 7.985 & 0.00831 & 37  &Sb    & Tspec/SOAR$^2$	& 1.1 & 2.0 & 40  \\
NGC 5128      & 13:25:27.6 & -43:01:08.8 & 3.942 & 0.00183 & 3.8 &S0$^0$& Tspec/SOAR$^2$	& 1.1 & 2.0 & 135  \\
ESO 021-G004  & 13:32:40.6 & -77:50:40.8 & 8.248 & 0.00984 & 39  &S0/a  & Tspec/SOAR$^2$	& 1.1 & 2.0 & 100  \\
MCG-06-30-015 & 13:35:53.7 & -34:17:44.1 & 9.582 & 0.00775 & 27  &E     & Tspec/SOAR$^2$	& 1.1 & 2.0 & 90  \\
NGC 5506      & 14:13:14.9 & -03:12:27.2 & 8.188 & 0.00608 & 27  &Sa    & Tspec/SOAR$^2$	& 1.1 & 2.0 & 90  \\
NGC 5728      & 14:42:23.9 & -17:15:11.4 & 8.171 & 0.00932 & 39  &Sa    & SpeX/IRTF$^5$	& 0.8 & 1.5 & 36 \\
ESO 137-G034  & 16:35:14.1 & -58:04:48.1 & 8.258 & 0.00914 & 35  &S0/a  & Tspec/SOAR$^2$	& 1.1 & 2.0 & 0  \\
NGC 6814      & 19:42:40.5 & -10:19:25.1 & 7.657 & 0.00522 & 23  &Sbc   & SpeX/IRTF$^7$	& 0.8 & 3.7 & 0  \\
NGC 7172      & 22:02:01.8 & -31:52:11.6 & 8.317 & 0.00868 & 37  &Sab   & Tspec/Blanco$^1$ & 1.1 & 2.0 & 97  \\
NGC 7213      & 22:09:16.2 & -47:10:00.0 & 7.035 & 0.00584 & 25  &Sa    & Tspec/Blanco$^1$ & 1.1 & 2.0 & 113 \\
NGC 7582      & 23:18:23.6 & -42:22:14.0 & 7.316 & 0.00525 & 22  &Sab   & Tspec/Blanco$^1$ & 1.1 & 2.0 & 119 \\
\hline
\noalign{\smallskip}
\multicolumn{9}{c}{Control Sample} \\
\noalign{\smallskip}
\hline
\noalign{\smallskip}
NGC 0718      & 01:53:13.3 & +04:11:45.3 & 8.739 & 0.00578 & 23  &Sa    & Tspec/Blanco$^1$ & 1.1 & 2.0 & 135  \\
NGC 1079      & 02:43:44.3 & -29:00:11.7 & 8.344 & 0.00484 & 19  &S0/a  & Tspec/SOAR$^2$	& 1.1 & 2.0 & 90   \\
NGC 1315      & 03:23:06.6 & -21:22:30.7 & 9.734 & 0.00539 & 21  &S0$^+$& Tspec/SOAR$^2$	& 1.1 & 2.0 & 90   \\
NGC 1947      & 05:26:47.6 & -63:45:36.0 & 7.497 & 0.00367 & 19  &S0$^-$& Tspec/SOAR$^2$	& 1.1 & 2.0 & 90   \\
ESO 208-G021  & 07:33:56.2 & -50:26:35.0 & 7.878 & 0.00362 & 17  &S0$^-$& Tspec/SOAR$^2$	& 1.1 & 2.0 & 110  \\
NGC 2775      & 09:10:20.1 & +07:02:16.5 & 7.037 & 0.00450 & 21  &Sab   & Tspec/SOAR$^2$	& 1.1 & 2.0 & 90   \\
NGC 3175      & 10:14:42.1 & -28:52:19.4 & 7.786 & 0.00363 & 14  &Sa    & Tspec/SOAR$^2$	& 1.1 & 2.0 & 45   \\
NGC 3351      & 10:43:57.7 & +11:42:13.0 & 6.665 & 0.00260 & 11  &Sb    & Tspec/SOAR$^2$	& 1.1 & 2.0 & 110  \\
ESO 093-G003  & 10:59:26.0 & -66:19:58.3 & 8.621 & 0.00611 & 22  &S0/a  & Tspec/SOAR$^2$	& 1.1 & 2.0 & 135  \\
NGC 3717      & 11:31:31.9 & -30:18:27.8 & 7.518 & 0.00578 & 24  &Sb    & Tspec/SOAR$^2$	& 1.1 & 2.0 & 90   \\
NGC 3749      & 11:35:53.2 & -37:59:50.3 & 8.705 & 0.00901 & 42  &Sa    & Tspec/SOAR$^2$	& 1.1 & 2.0 & 120  \\
NGC 4224      & 12:16:33.7 & +07:27:43.6 & 8.609 & 0.00862 & 41  &Sa    & Tspec/SOAR$^2$	& 1.1 & 2.0 & 60   \\
NGC 4254      & 12:18:49.6 & +14:24:59.3 & 6.929 & 0.00803 & 15  &Sc    & Tspec/SOAR$^2$	& 1.1 & 2.0 & 90   \\
NGC 4260      & 12:19:22.2 & +06:05:55.6 & 8.538 & 0.00592 & 31  &Sa    & Tspec/SOAR$^2$	& 1.1 & 2.0 & 50   \\
NGC 5037      & 13:14:59.3 & -16:35:25.1 & 8.599 & 0.00619 & 35  &Sa    & Tspec/SOAR$^2$	& 1.1 & 2.0 & 45   \\
NGC 5845      & 15:06:00.7 & +01:38:01.7 & 9.112 & 0.00555 & 25  &E     & Tspec/SOAR$^2$	& 1.1 & 2.0 & 135  \\
NGC 5921      & 15:21:56.5 & +05:04:13.9 & 8.096 & 0.00494 & 21  &Sbc   & Tspec/SOAR$^2$	& 1.1 & 2.0 & 20   \\
IC 4653       & 17:27:06.9 & -60:52:44.1 & 10.08 & 0.00640 & 26  &S0/a  & Tspec/SOAR$^2$	& 1.1 & 2.0 & 131  \\
NGC 7727      & 23:39:53.8 & -12:17:34.8 & 7.688 & 0.00599 & 26  &Sa    & Tspec/Blanco$^1$ & 1.1 & 2.0 & 10   \\
\hline
\end{tabular}
	\begin{list}{Table Notes:}
	\item(1) Galaxy name; (2) Right Ascension; (3) Declination; (4) K band magnitude \citep{Jarrett+03}; (5) Redshift taken from NED (https://ned.ipac.caltech.edu/); (6) Distance from \citet{Davies+15}; (7) Morphology, as compiled by \citet{Davies+15} and converted to \citet{DeVaucouleurs+59} system; (8) Spectrograph and telescope used; (9) Slit width of the instrument; (10) Aperture integrated for final spectra; (11) Position angle of the slit, in degrees East of North.
    \item $^1$ Obtained under project 2016B-0912; $^2$ Obtained under project SO2021B-004; $^3$ Obtained under project 2017A022; $^4$ Obtained under project 2016A-0621; $^5$ Data published by \citet{Riffel+06}; $^6$ Data published by \citet{Mason+15}; $^7$ Data published by \citet{Riffel+19}.
	\end{list}

\end{table*}

\begin{figure*}
\begin{minipage}[b]{0.5\linewidth}
\includegraphics[width=\textwidth]{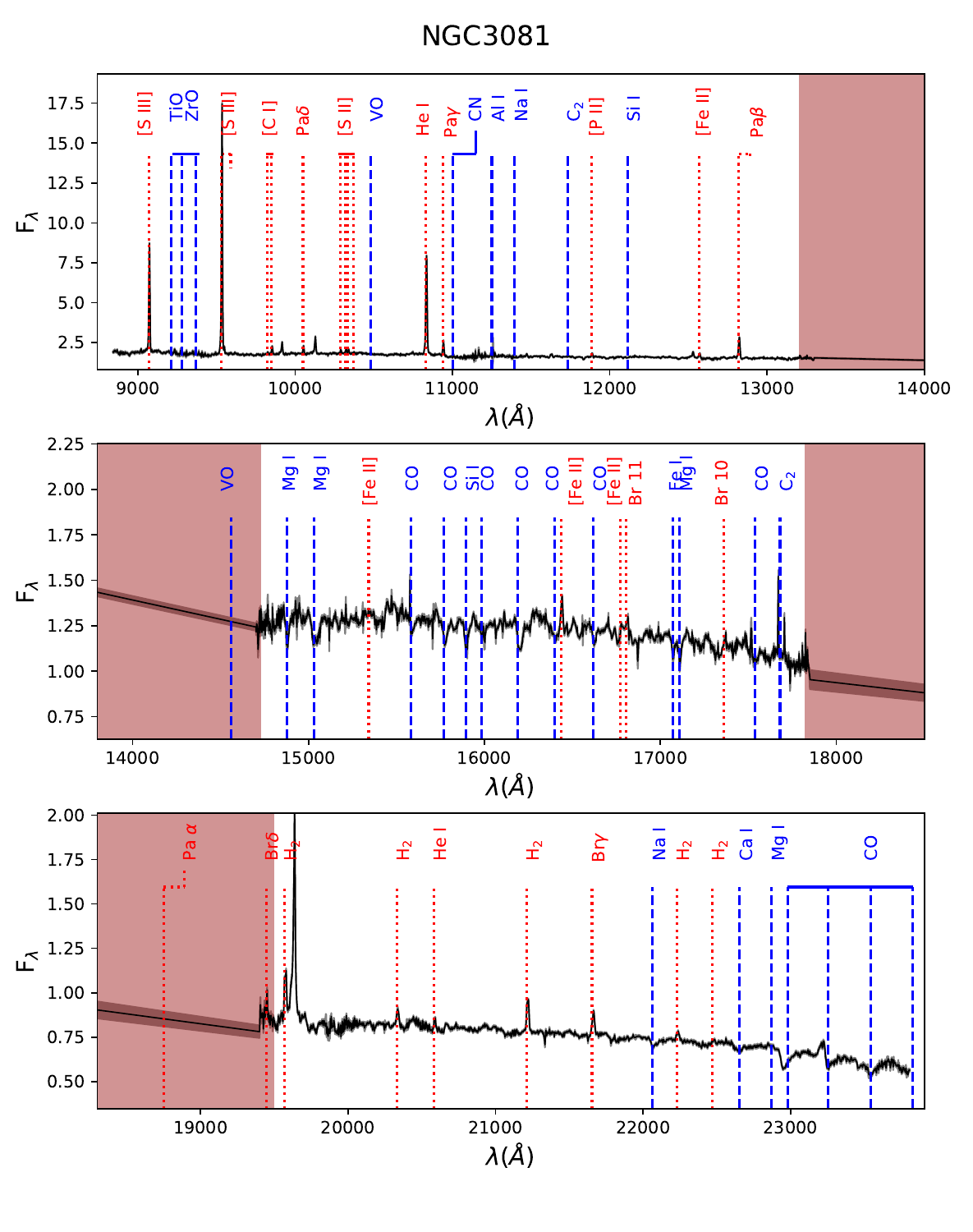}
\end{minipage}\hfill
\begin{minipage}[b]{0.5\linewidth}
\includegraphics[width=\textwidth]{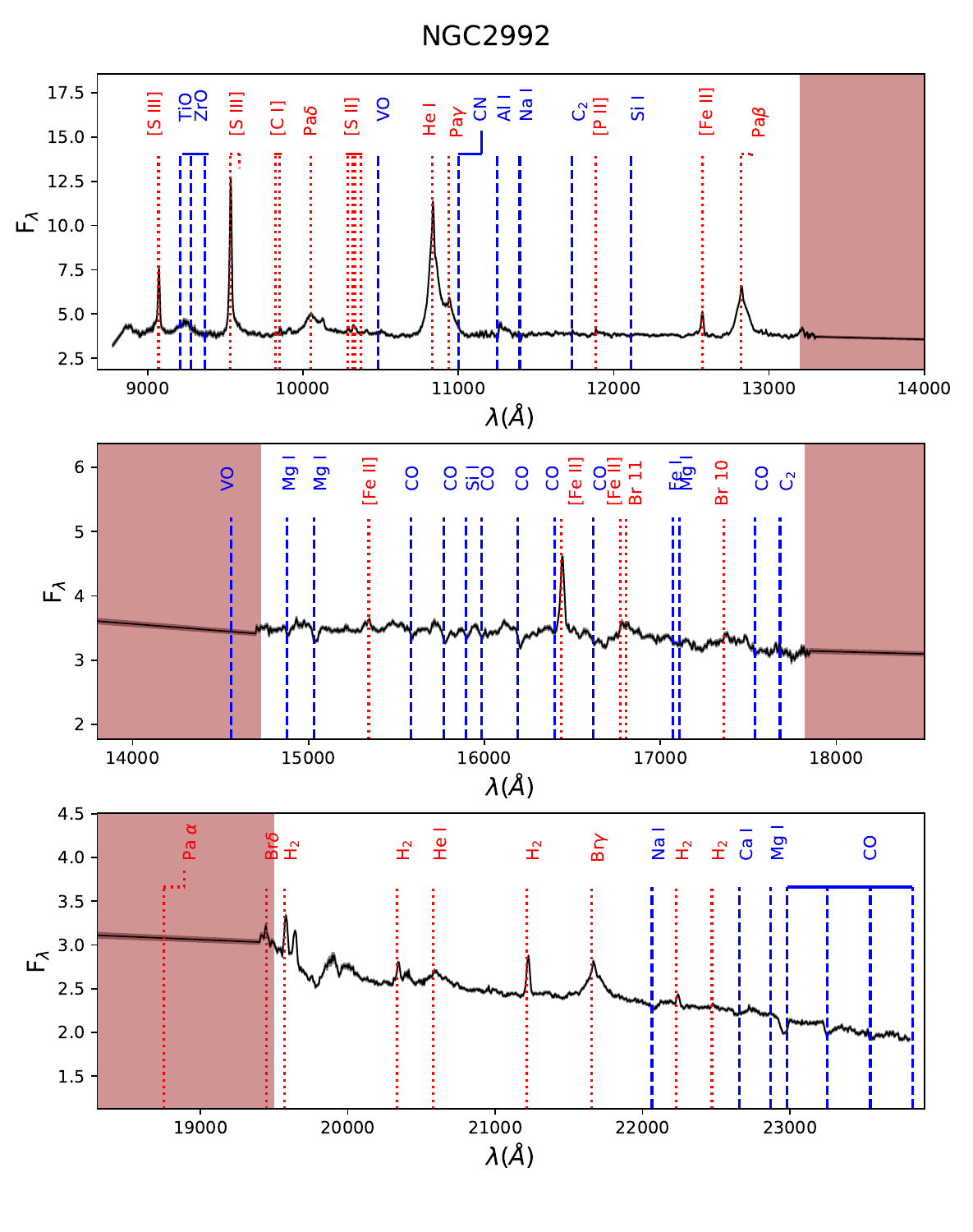}
\end{minipage}\hfill
\begin{minipage}[b]{0.5\linewidth}
\includegraphics[width=\textwidth]{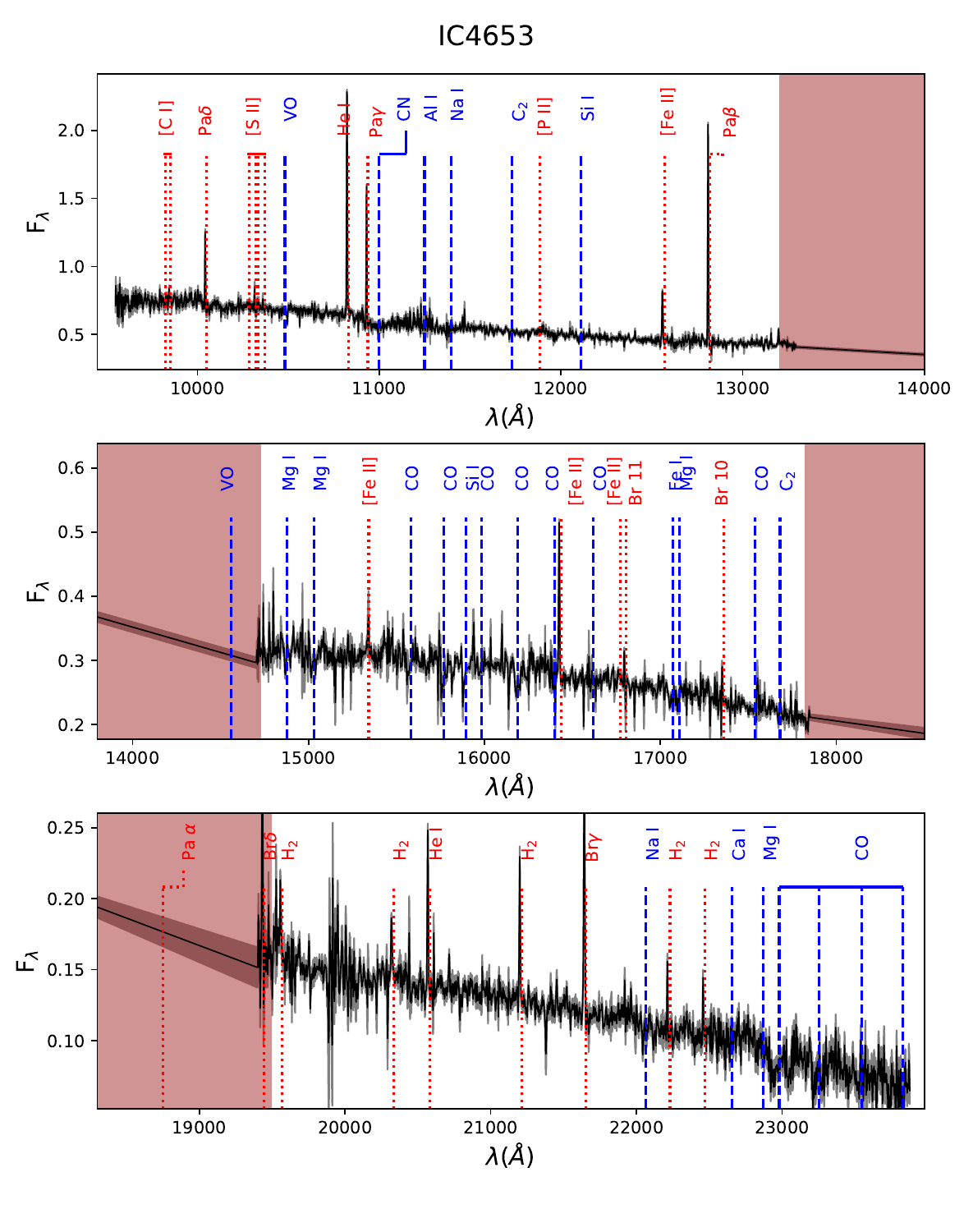}
\end{minipage}\hfill
\begin{minipage}[b]{0.5\linewidth}
\includegraphics[width=\textwidth]{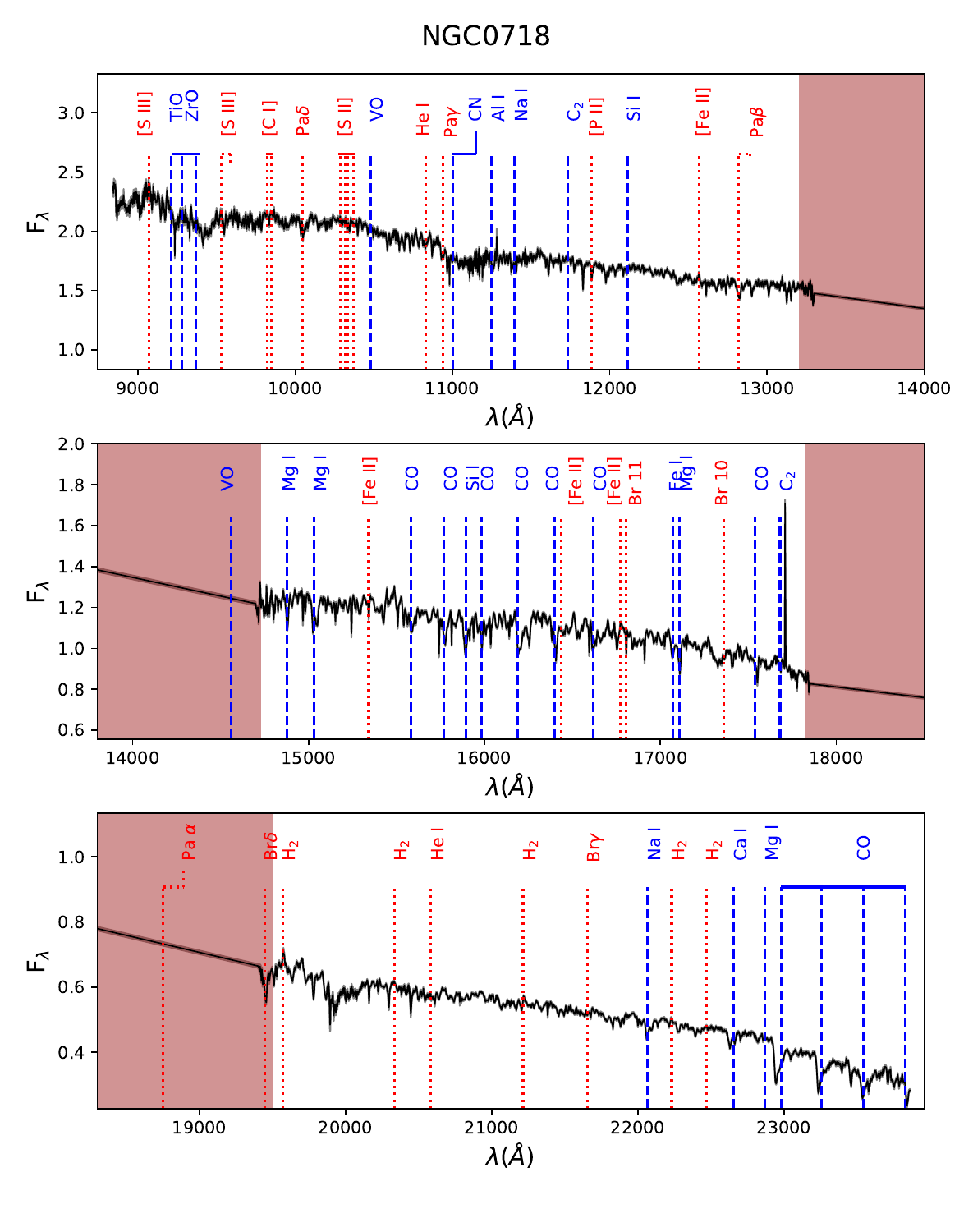}
\end{minipage}\hfill
\caption{Final reduced and redshift-corrected spectra for NGC~3081 (AGN), NGC~2992 (AGN), IC~4653 (CNT) and NGC~0718 (CNT). For each galaxy we show -- the redshift and reddening corrected spectra-- from top to bottom the, $J$, $H$ , and $K$ bands, respectively. The flux is in units of  $\rm 10^{-15}~ erg ~ cm^{-2} ~ s^{-1}$. The shaded grey area represents the uncertainties and the brown area indicates the poor transmission regions between different bands. The remaining spectra are shown in on-line material.}
\label{spectra}
\end{figure*}

After the data reduction, we corrected our spectra for galactic reddening using a \citet[][hereafter CCM]{Cardelli+89} law, and then corrected for redshift (this was done also for the literature data) and homogenized the spectral resolution of them. An example of the final reduced spectra for four galaxies, namely: NGC~3081 (AGN), NGC~2992 (AGN), IC~4653 (control galaxy, hereafter CNT), and NGC~0718 (CNT) are shown in Fig.~\ref{spectra}, together with the identification of the most prominent absorption- and emission-line features. The remaining sources are shown as Supplementary Material. Observing information and some general properties of the galaxies are shown in Table~\ref{tab:obs_log} with a full description of the sample made in \citet{Davies+15}. The main AGN properties are shown in Table~\ref{tab:AGN}.

\begin{table*}
\centering
\caption{Main AGN properties.}
\label{tab:AGN}
\begin{tabular}{llcccccccc}
\hline
Source        &Activity   & L$_{X}^{obs}$ & L$_{X}^{int}$ & M$_{SMBH}$           & log~$L_H$ & log N$_H$  & $log(L'_{CO})$           & log~$L_{[O III]}$ & log~$L_{H_2}$     \\ 
              &~~Type     &  erg~s$^{-1}$ & erg~s$^{-1}$  &10$^6M_\odot$         & L$_\odot$ & cm$^{-2}$  & K km s$^{-1}$\,$pc^2$    & erg~s$^{-1}$      &  erg~s$^{-1}$     \\ 
~~~~(1)       &~~~(2)     & (3)           &    (4)        &  (5)                 & (6)       &    (7)    &    (8) 			     &   (9)             &   (10)            \\ 
\hline
ESO021-G004   &Sy 2       & 42.56         & 42.70         & 52.1$\pm$38.4$^3$    & 10.53     & 23.8 	  & 8.083                    & 38.68$\pm$0.46    &  	  --		 \\
ESO137-G034   &Sy 2       & 42.71         & 42.76         & 21.5$\pm$15.8$^3$    & 10.44     & 24.3 	  & 7.820                    & 40.66$\pm$0.45    &  38.81$\pm$0.45   \\
MCG-05-14-012 &Sy 1.0     & 42.64         & 42.65         & 2.29$\pm$0.68$^1$    & 9.60      & $\leq$21.9 & --                       & 39.21$\pm$0.46    &  	 -- 		 \\
MCG-05-23-016 &Sy 1.9     & 43.51         & 43.50         & 27.1$\pm$8.74$^1$    & 9.94      & 22.2 	  & 7.445                    & 39.62$\pm$0.46    &  38.31$\pm$0.45   \\
MCG-06-30-015 &Sy 1.2     & 42.92         & 42.91         & 7.38$\pm$1.98$^1$    & 9.59      & 20.9 	  & 7.109                    & 39.65$\pm$0.47    &  37.62$\pm$0.45   \\
NGC1365       &Sy 1.8     & 42.63         & 42.60         & 19.7$\pm$5.77$^1$    & 10.58     & 22.2 	  & 8.782                    &	  --		     &  38.12$\pm$0.45   \\
NGC2110       &Sy 2       & 43.63         & 43.63         &  150$\pm$110$^3$     & 10.44     & 23.0 	  & 7.603                    & 40.15$\pm$0.45    &  39.11$\pm$0.46   \\
NGC2992       &Sy 1.8     & 42.55         & 42.52         & 22.8$\pm$6.74$^1$    & 10.31     & 21.7 	  & 8.472                    & 39.99$\pm$0.45    &  39.26$\pm$0.46   \\
NGC3081       &Sy 2       & 43.07         & 43.29         & 36.6$\pm$26.9$^3$    & 10.15     & 23.9 	  & 7.749                    & 40.57$\pm$0.45    &  38.73$\pm$0.45   \\
NGC3783       &Sy 1.2     & 43.58         & 43.58         & 11.2$\pm$3.61$^1$    & 10.29     & 20.5 	  & 7.955                    & 40.76$\pm$0.45    &  38.02$\pm$0.45   \\
NGC4235       &Sy 1.2     & 42.66         & 42.64         & 55.8$\pm$15.9$^1$    & 10.43     & 21.3 	  & 7.754                    & 39.39$\pm$0.45    &  	 -- 		 \\
NGC4388       &Sy 2       & 43.64         & 43.70         & 8.40$\pm$0.2$^2$     & 10.65     & 23.5 	  & 8.151                    &	  --		     &  39.00$\pm$0.46   \\
NGC4593       &Sy 1.0/1.2 & 43.20         & 43.20         & 12.4$\pm$3.91$^1$    & 10.59     & $\leq$19.2 & 8.146                    & 39.70$\pm$0.45    &  	  --		 \\
NGC5128       &Sy 2       & 43.00         & 43.02         & 66.3$\pm$48.9$^3$    & 10.22     & 23.1 	  & --                       &	  --		     &  37.12$\pm$0.46   \\
NGC5506       &Sy 2       & 43.31         & 43.30         & 22.4$\pm$17.2$^3$    & 10.09     & 22.4 	  & 7.874                    &	  --		     &  38.74$\pm$0.45   \\
NGC5728       &Sy 2       & 43.23         & 43.36         & 23.0$\pm$2.3$^2$     & 10.56     & 24.2 	  & 8.531                    & 40.50$\pm$0.45    &  39.04$\pm$0.46   \\
NGC6814       &Sy 1.5     & 42.76         & 42.75         & 11.6$\pm$3.67$^1$    & 10.31     & 21.0 	  & 7.491                    & 39.57$\pm$0.45    &  37.73$\pm$0.45   \\
NGC7172       &Sy 2       & 43.46         & 43.32         & 53.4$\pm$39.3$^3$    & 10.43     & 22.9 	  & 8.658                    & 38.60$\pm$0.45    &  	--  		 \\
NGC7213       &Sy 1       & 42.50         & 42.49         & 6.46$\pm$2.01$^1$    & 10.62     & $\leq$20.4 & 7.959                    & 39.26$\pm$0.45    &  38.47$\pm$0.45   \\
NGC7582       &Sy 2       & 42.68         & 43.29         & 30.5$\pm$22.4$^3$    & 10.38     & 24.2 	  & 8.917                    & 39.29$\pm$0.45    &  38.38$\pm$0.45   \\
\hline
\end{tabular}
\begin{list}{Table Notes:}
\item (1) Source name; 
(2) Activity type, as compiled by \citet{Davies+15}, with the exception of ESO021-G004, whose classification was taken from \citet{Burtscher+21}; 
(3) Logarithm of the observed hard X-ray luminosity (14-195~keV) from \citet{Ricci+17}; 
(4) Logarithm of the intrinsic hard X-ray luminosity (14-195~keV) from \citet{Ricci+17};
(5) The Black Hole mass, published by \citet{Caglar+20}, derived from: $^1$broad H$\alpha$ emission, $^2$megamaser or $^3$stellar velocity dispersion;
(6) Integrated H-band luminosity \citep[given by the 2MASS total magnitude][]{Skrutskie+06}; 
(7) Neutral absorbing column, from \citet{Ricci+15}, based on modeling 0.3-150~keV spectrum; 
(8) The CO luminosities, from \citet{Rosario+18}; 
(9) Measured [\ion{O}{iii}] luminosity, and
(10) Measured \h2\ luminosity.
\end{list}
\end{table*}

\section{Stellar Population Fitting}\label{popstar}

The integrated spectra of galaxies are comprised of various components, including for example the underlying stellar, gas, and dust emission contributions \citep[][and references therein]{Bica+87,Schmidt+91,Walcher+11,Conroy+13,Gomes+17,Riffel+09,Riffel+11c,Riffel+22}. In the case of active galaxies, additional considerations must be given to components such as the AGN torus and accretion disk \citep{Riffel+09,Burtscher+15,Riffel+22}. The process of stellar population fitting involves determining the percentage contribution of these components to the integrated spectrum.

It is important to note that due to the large number of parameters involved (such as age, metallicity, kinematics, reddening, and AGN components), numerous techniques have been employed over the years to separate the different components of a galaxy's spectrum \citep[see][for a review]{Walcher+11,Conroy+13}. As a result, various fitting codes have been developed by different research groups, each with their own priorities in mind \citep[e.g.][]{CidFernandes+05,Ocvirk+06,Koleva+09,Tojeiro+07,Cappellari+17,Sanchez+16,Wilkinson+17,Gomes+17,Johnson+21}. Naturally, comparisons among these different codes have been conducted, demonstrating that, in general, the codes yield consistent results when applied to the same data with the same input parameters \citep[e.g.][]{Koleva+08,Dias+10,Gomes+17,Goddard+17,Ge+18,CidFernandes+18,Woo+24}.

For the purpose of conducting stellar population fitting, we use the software \textsc{starlight} \citep{CidFernandes+04,CidFernandes+05,Asari+07,CidFernandes+18}. We opted for this particular choice primarily to ensure consistency with previous studies conducted by our team, enabling easier comparisons of the results. The \textsc{starlight} code fits the complete absorption and continuum features in the observed spectra by combining in different proportions the base-set elements. It excludes emission lines and spurious data, employing a blend of computational techniques derived from semi-empirical population synthesis and evolutionary synthesis models \citep{CidFernandes+04,CidFernandes+05}.

In essence, the code fits an observed spectrum, represented as $O_{\lambda}$, using a combination of $N_{\star}$ simple stellar populations (SSPs) in varying proportions. The visual extinction ($A_V$) is modeled by \str\ as foreground dust lanes.
In the fits we use the CCM \citep{Cardelli+89} extinction law. 
The modeled spectrum, $M_{\lambda}$, is obtained through the following equation:

\begin{equation}
M_{\lambda}=M_{\lambda 0}\sum_{j=1}^{N_{\star}}x_j\,b_{j,\lambda}\,r_{\lambda}  \otimes G(v_{\star},\sigma_{\star})
\label{streq}
\end{equation}
where $x_j$ is the population vector, $b_{j,\lambda}$ is the $j$th base element (see below), $\,r_{\lambda}$ is the 
reddening factor of the $j$th component normalised at $\lambda_0$, the reddening term is represented by 
$r_{\lambda}=10^{-0.4(A_{\lambda}-A_{\lambda 0})}$, $M_{\lambda 0}$ is the
synthetic flux at the normalisation wavelength (we have used $\rm \lambda_{norm}=12\,230$\AA\ in the rest frame).  The convolution 
operator is $\otimes$ and $G(v_{\star},\sigma_{\star})$ is the Gaussian distribution used to model the line-of-sight velocity distributions of the stars, which is centered at velocity $v_{\star}$  with dispersion  $\sigma_{\star}$. 
The final fit is carried out through a chi-square minimisation, as follows: 

\begin{equation}
\chi^2 = \sum_{\lambda}[(O_{\lambda}-M_{\lambda})w_{\lambda}]^2
\end{equation}
where emission lines and spurious features are excluded from the fit by fixing $w_{\lambda}$=0  at the corresponding wavelengths. 

For the base of elements, in the present paper, we used as simple stellar population (SSP) the evolutionary population synthesis (EPS) models of \citet{Verro+22b}. These models have been computed with the new X-shooter Spectral Library \citep[XSL,][]{Verro+22}. The base of elements comprises SSPs with 4 metalicities ($Z$ = 0.25, 0.63, 1 and 1.53 $Z_\odot$) and 25 ages (t = 0.050, 0.063, 0.079, 0.1, 0.126, 0.158, 0.200, 0.251, 0.316, 0.398, 0.501, 0.631, 0.794, 1, 1.259, 1.585, 1.995, 2.512, 3.162, 3.981, 5.012, 6.310, 7.943, 10 and, 12.589~Gyr). They have been selected to have \citet{Kroupa+01} IMF and the {\sc parsec/colibri} isochrones \citep{Chen+15,Pastorelli+20}, cover younger ages and are on the models safe-range \citep[see Fig. 19 of][]{Verro+22b}.

\begin{figure}
    \centering
    \includegraphics[scale=0.5]{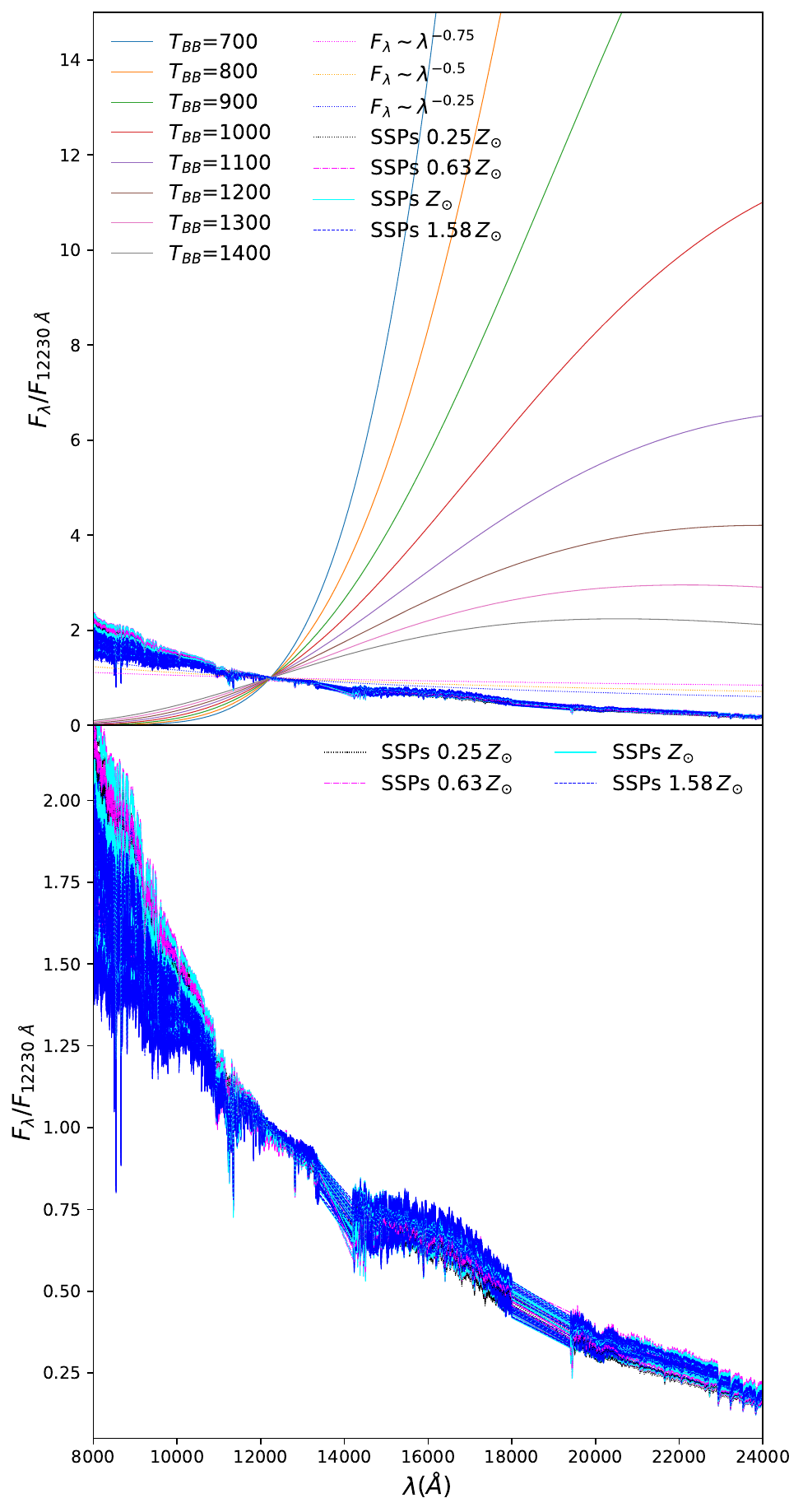}
    \caption{The base of elements comprises SSPs with 4 metalicities ($Z$ = 0.25, 0.63, 1 and 1.53 $Z_\odot$) and 25 ages (t = 0.050, 0.063, 0.079, 0.1, 0.126, 0.158, 0.200, 0.251, 0.316, 0.398, 0.501, 0.631, 0.794, 1, 1.259, 1.585, 1.995, 2.512, 3.162, 3.981, 5.012, 6.310, 7.943, 10 and, 12.589~Gyr). The accretion disk (FC) is described as a power-law of the form $F_\lambda \propto \lambda^{-\alpha}$ (we used  $\alpha$=0.25, 0.5, and 0.75). Hot dust emission  component are described as eight Planck distributions
(black-body, BB), with temperature ranging from 700 to 1400 K, in steps of 100 K. For display purposes on the bottom plot we zoom in the SSPs. The identification of the components are on the labels. For details see text.}
    \label{base}
\end{figure}

Additionally, to account for the accretion disk featureless continuum (FC)  we have used a power-law of the form $F_\lambda \propto \lambda^{-\alpha}$ \citep[e.g.][]{Koski+78,CidFernandes+05,Riffel+09}. We employed three different values for $\alpha$, namely: 0.25, 0.5, and 0.75.  To properly account for the hot dust emission  component, eight Planck distributions
(black-body, BB), with temperatures ranging from 700 to 1400 K, in steps of 100 K, were included in the fits. The lower limit was chosen because lower temperatures are hard to detect in this spectral range \citep{Riffel+09} and the upper limit is very close to the sublimation temperature of the dust grains \citep[e.g.][]{Barvainis+87,Granato+94}.  For more details on the effects of these components in the NIR spectra and on the definition of our base of elements see \citet{Riffel+09} and \citet{Riffel+22}. The components used in the fits are shown in Fig.~\ref{base}.

\subsection{Uncertainties on the fits} \label{error}

Statistical analysis plays a crucial role in interpreting spectral fitting findings, as evident from its widespread application \citep[e.g.][]{Panter+07}. The practice of averaging results serves to diminish uncertainties \citep{CidFernandes+13}. Incorporating statistical interpretation is also advantageous for generating uncertainty estimations, a feature not readily available in \str\ standard output.

A viable approach to obtain such estimations involves perturbing the input spectra, considering realistic error prescriptions \citep{CidFernandes+13}. Here we followed \citet{Dametto+14} and simulated 100 spectra for each galaxy within our sample employing Monte Carlo approach.  The simulated flux for each wavelength ($\lambda_i$) is computed assuming a Gaussian distribution of the uncertainties, being the mean flux in each $\lambda_i$ the measured flux ($F_{\lambda_i}$) value, and the standard deviation is given by the errors derived in the data reduction (which include the propagation of uncertainties over all the data reduction steps). 

During its likelihood-guided sampling process of the parameter space, \str\ offers a single best-fitting set of parameters from the numerous trials during the fitting process. Nevertheless, it is important to note that these individual solutions may not always be mathematically unique. In light of this, we proceeded with stellar population fitting on all the simulated spectra, aiming to derive an average value for each galaxy, along with the associated standard deviation. This approach allows us to gain a more robust and comprehensive understanding of the results, considering the inherent uncertainties present in the analysis. 


\section{Results}\label{sec:results}
\subsection{Fitting Approach}
\begin{figure*}
\begin{minipage}[b]{0.48\linewidth}
\includegraphics[width=\textwidth]{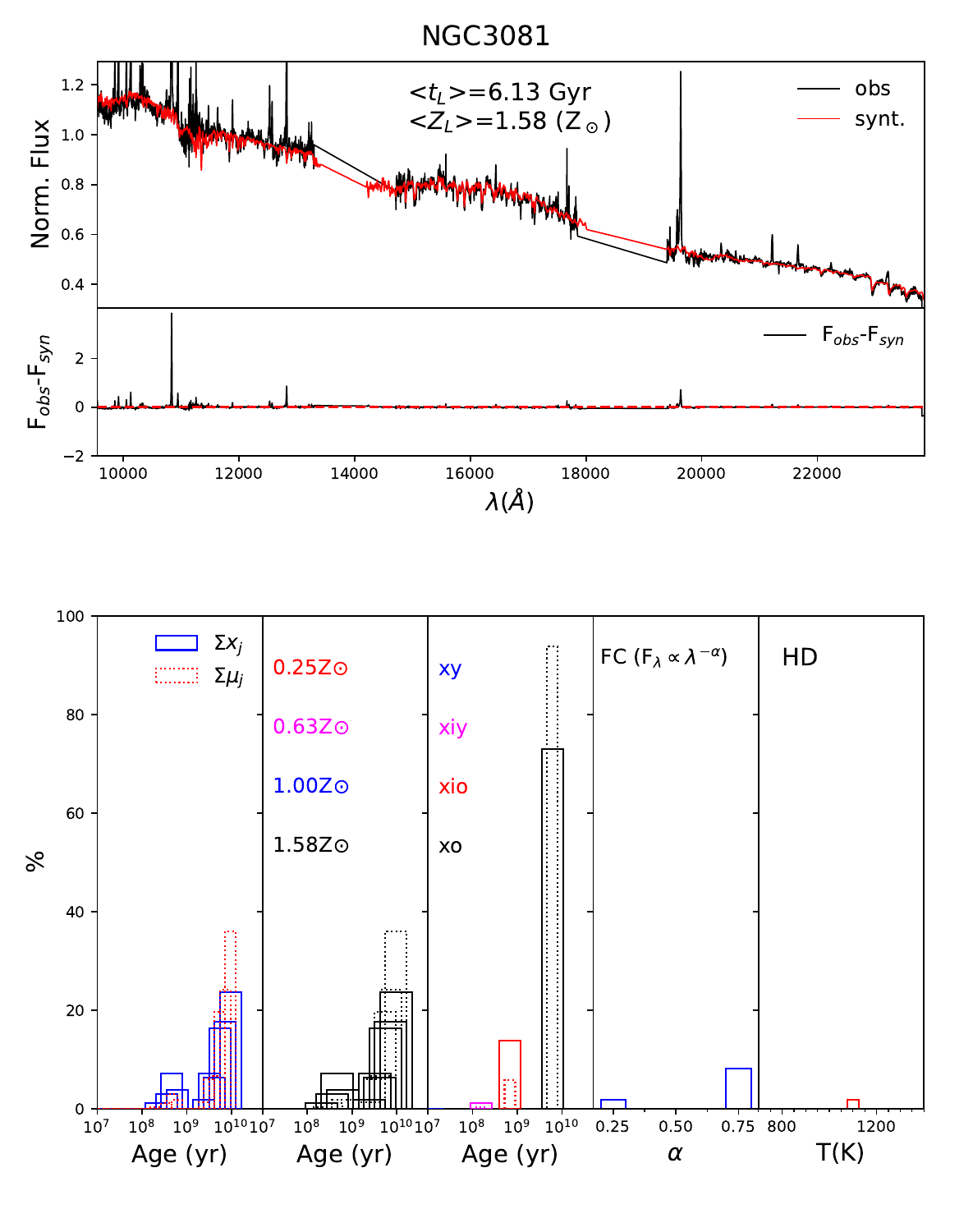}
\end{minipage}\hfill
\begin{minipage}[b]{0.48\linewidth}
\includegraphics[width=\textwidth]{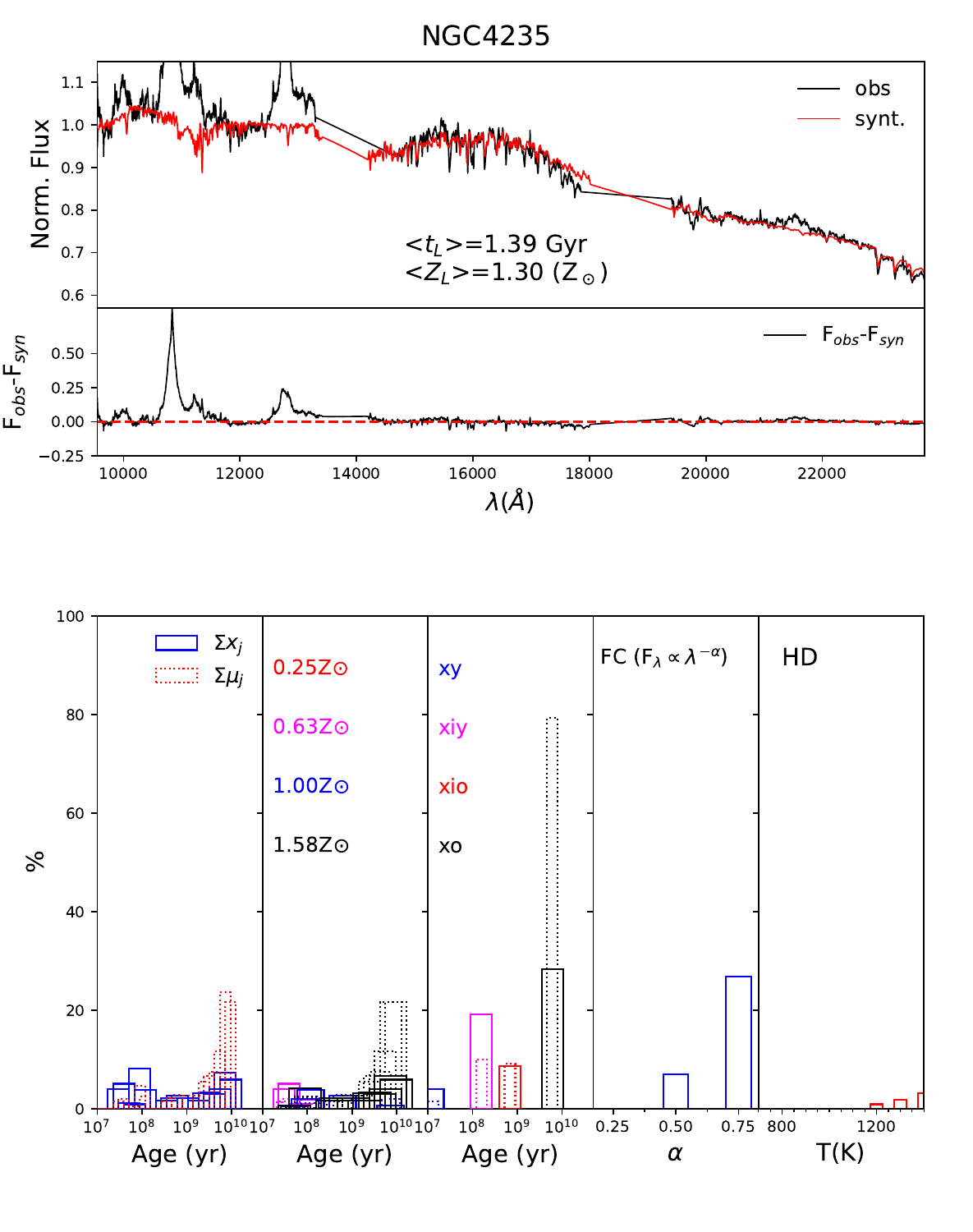}
\end{minipage}\hfill
\begin{minipage}[b]{0.48\linewidth}
\includegraphics[width=\textwidth]{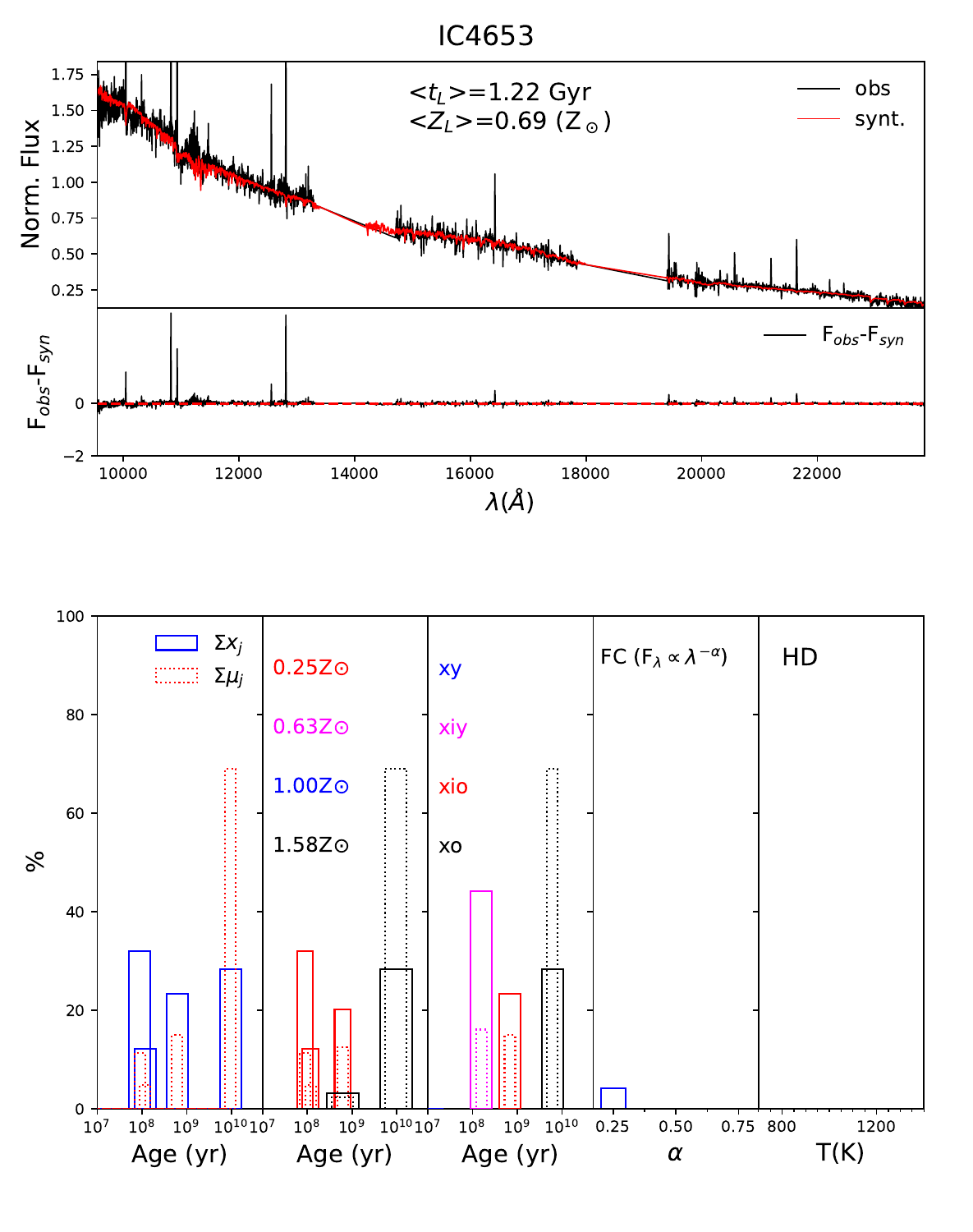}
\end{minipage}\hfill
\begin{minipage}[b]{0.48\linewidth}
\includegraphics[width=\textwidth]{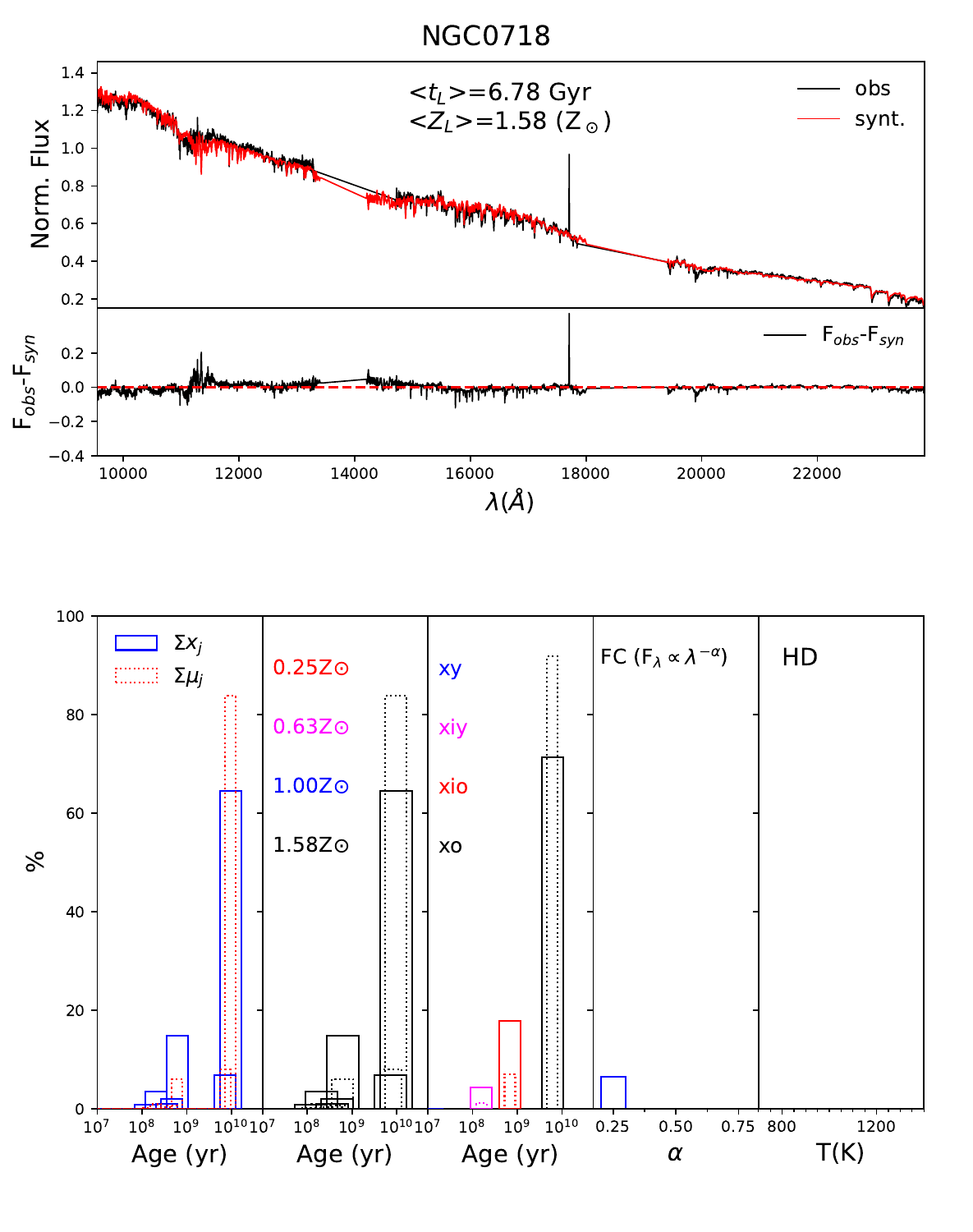}
\end{minipage}\hfill
\caption{Spectral fitting results for NGC~3081 (AGN, Sy~2), NGC~4235 (AGN, Sy~1), IC~4653 (CNT, star-forming) and NGC~0718 (CNT, passive) for the original spectrum (e.g. without the perturbations described in \S~\ref{error}). For {\bf each galaxy}, on 
{\it top panel}: observed (black) and synthetic (red) spectra, light-weighted mean age, and metallicity.
{\it middle panel}: emission-line free spectrum (observed$-$synthetic spectrum) and a horizontal line showing the zero point for the continuum. 
{\it bottom panel}: from left to right, SFH summed over all metalicities in light ($x$) and in mass ($\mu$) fractions; SFH considering different metallicities (they are labelled); binned population vectors; AGN feature-less continuum and hot-dust contributions are on the two last panels. The plots for the remaining sources are shown in on-line material.}
\label{sp_fits}
\end{figure*}

The results of the stellar population fitting are the fractional contributions, in light and mass (in the case of the stellar ones), of each one of the base components (the so-called population vector). They can be separated into stellar population components and AGN components.
An example of the results of the fitting, the population vectors, binned population vectors (see below), as well as mean ages and metallicities are shown in Fig.~\ref{sp_fits}, where we show the fits for four sources, namely NGC~3081, NGC~4235, IC~4653, and NGC~0718, two AGN hosts and two control sources, respectively. For each galaxy, we show on the {\it upper top panel} the observed spectrum (black) and the synthetic spectrum (red); in the {\it middle panel} the "pure" emission-spectrum (difference between the observed and synthetic spectrum). On the {\it bottom panel}: from left to right, SFH summed over all metallicities in light ($x$) and mass ($\mu$) fractions; SFH considering different metallicities (they are labelled); binned population vectors (see below); AGN FC continuum and HD contributions are on the two last panels. The plots for the remaining spectra are shown in the supplementary material.

To consider the impact of noise on the data, which can obscure differences between similar spectral components, as well as the age-metallicity degeneracy \citep[e.g.][]{Worthey+94}, besides the approach presented in \S~\ref{error} we also adopted the approach used by \citet{CidFernandes+05} and used the {\it binned} (or condensed) population vectors\footnote{By binning the data, we obtain coarser but more reliable fractions for each bin, thereby accounting for uncertainties arising from noise effects and spectral similarities.} as defined in \citet{Riffel+09,Riffel+22}:

\begin{description}
    \item[$xy$:] sum of the percent  contribution of SSPs with ages 50 $\leq t \leq$ 100~Myr. This vector is representative of the younger (ionizing) population\footnote{Note however that it is limited by the younger age available in the XSL models.}.
           
    \item[$xiy$:] sum of the percent contribution of SSPs with ages in the range 100~Myr $< t \leq$ 700~Myr. This represents the peak of the Balmer absorption lines. 
    
    \item[$xio$:] sum of the percent contribution SSPs with ages in the range 700~Myr $< t \leq$ 2~Gyr. Being the representative of the AGB peak contribution. 
    
    \item[$xo$:] sum of the percent contribution 2~Gyr $< t \leq$ 13~Gyr. Represents the older underlying population. 
    
    \item[HD:]  sum of the percent contribution of all Planck function components, representing hot dust emission.
    
    \item[FC:] sum of the percent contribution of all power-law featureless components, representing the AGN accretion disc contribution.
 \end{description}

If limited to selecting only two parameters to describe the stellar population composition of a galaxy, the optimal choice would undoubtedly be its mean age and mean-metallicity. Therefore, we have followed \citet[][]{CidFernandes+05} and computed the mean ages (the logarithm of the age, actually) for each galaxy weighted by the stellar light 
\begin{equation}\langle {\rm log} t_{\star} \rangle_{L} = \displaystyle \sum^{N_{\star}}_{j=1} x_j {\rm log}t_j,  \label{magel} \end{equation}
and weighted by the stellar mass, 
\begin{equation}\langle {\rm log} t_{\star} \rangle_{M} = \displaystyle \sum^{N_{\star}}_{j=1} \mu_j {\rm log}t_j. \label{magem} \end{equation}
Subsequently, we convert these quantities into ages in Gyr (e.g., removing the logarithmic scale).

The light-weighted mean metallicity is defined as  
\begin{equation}\langle Z_{\star} \rangle_{L} = \displaystyle \sum^{N_{\star}}_{j=1} x_j Z_j, \label{mzl}\end{equation}
and the mass-weighted mean metallicity is defined by:
\begin{equation}\langle Z_{\star} \rangle_{M} = \displaystyle \sum^{N_{\star}}_{j=1} \mu_j Z_j. \label{mzm}\end{equation}

Note that all these values are limited by the elements included in the base. In principle, the mass-weighted properties are more physical, but, since the stellar $M/L$ is non-constant, they have a much less direct relation with the observed spectrum than the light-weighted ones. For more details see \citet{CidFernandes+05}.

Besides the actual stellar mass of the galaxy\footnote{Note that this is the mass inside the aperture size, see \S~\ref{stelMass} for a discussion.} ($M_{\star}$), \str\ also outputs the mass that has been processed into stars throughout the galaxy's life ($M_{\star}^{ini}$). Using this, one can compute the mass that has returned to the interstellar medium, as $M_{\star}^{Ret}$ = $M_{\star}^{ini}$ - $M_{\star}$. These quantities are listed in Tab.~\ref{tab:spfits}.

Finally, these quantities have been derived as the mean value of all the fitting on the simulated spectra with the uncertainty being its standard deviation (see \S~\ref{error}). 

\subsection{Stellar population properties}

\begin{figure*}
\includegraphics[width=\textwidth]{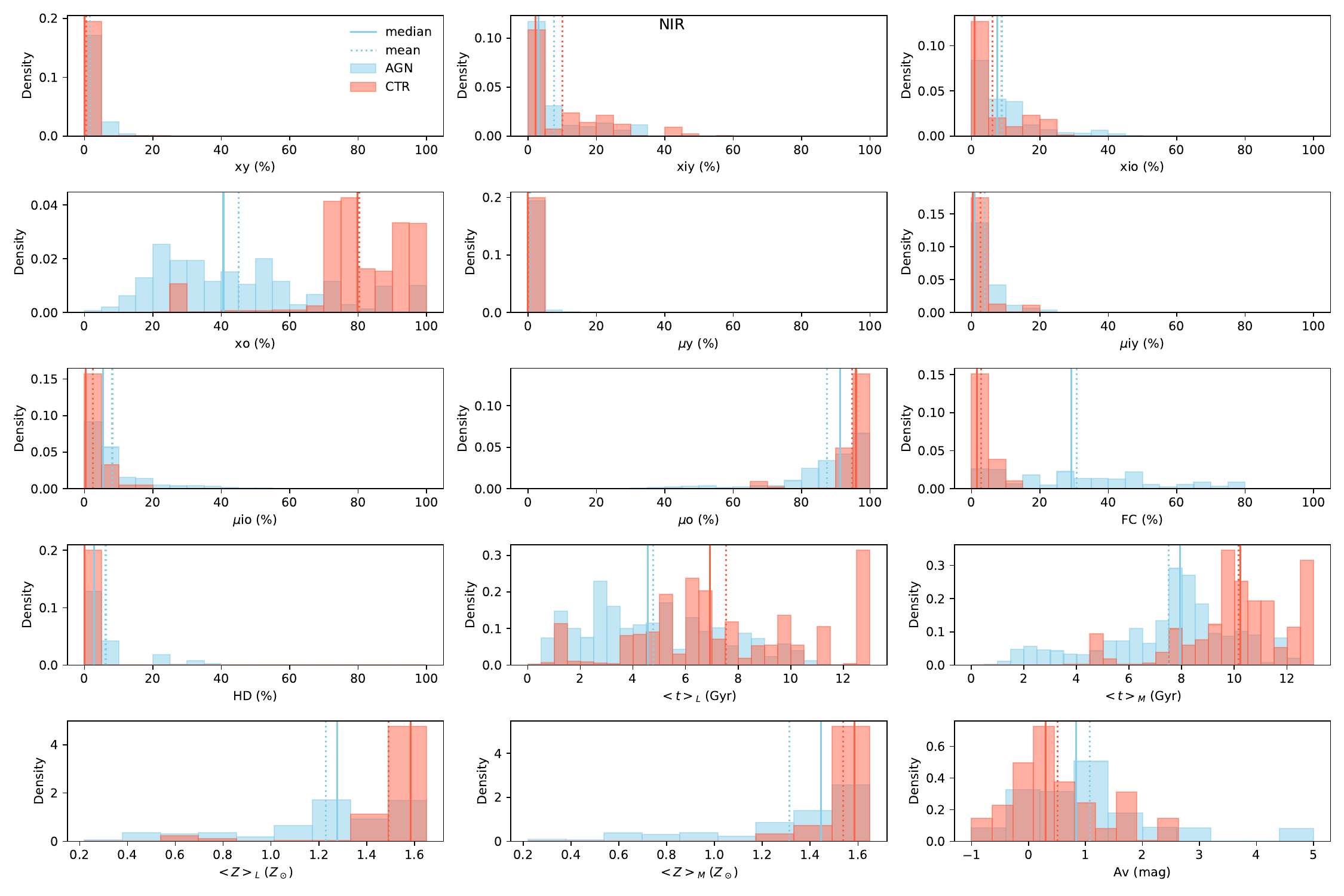}
\caption{Mean stellar population properties of the sample obtained after fitting the 100 simulated spectra (see \S~\ref{error}). In blue AGN hosts and red control galaxies. Mean (dotted line) and median (full line) are shown. Population vectors are defined as follows: $xy\leq$ 100~Myr, 100~Myr $< xiy \leq$ 700~Myr,  700~Gyr $< xio \leq$ 2~Gyr and 2~Gyr $< xo \leq$ 13~Gyr. HD and FC are the sum of the percent contribution of the Planck function and the power-law featureless components, respectively. Mean age and metallicities are defined according to equations:~\ref{magel},~\ref{magem},~\ref{mzl} and \ref{mzm}. The mean values and their standard deviations are listed in Tab.~\ref{tab:spfits}.}
\label{fig:NIRMeanProps}
\end{figure*}

In Fig.~\ref{fig:NIRMeanProps} we show the distribution (mean and median values are also shown) of the results of the 100 simulated spectra for each galaxy (see \S~\ref{error}) for the binned population vectors as well as for mean ages and metallicities for the AGN hosts (blue) and control galaxies (red). The mean values and their standard deviations are summarised in Tab.~\ref{tab:spfits}.

\begin{figure*}
\includegraphics[width=0.9\textwidth]{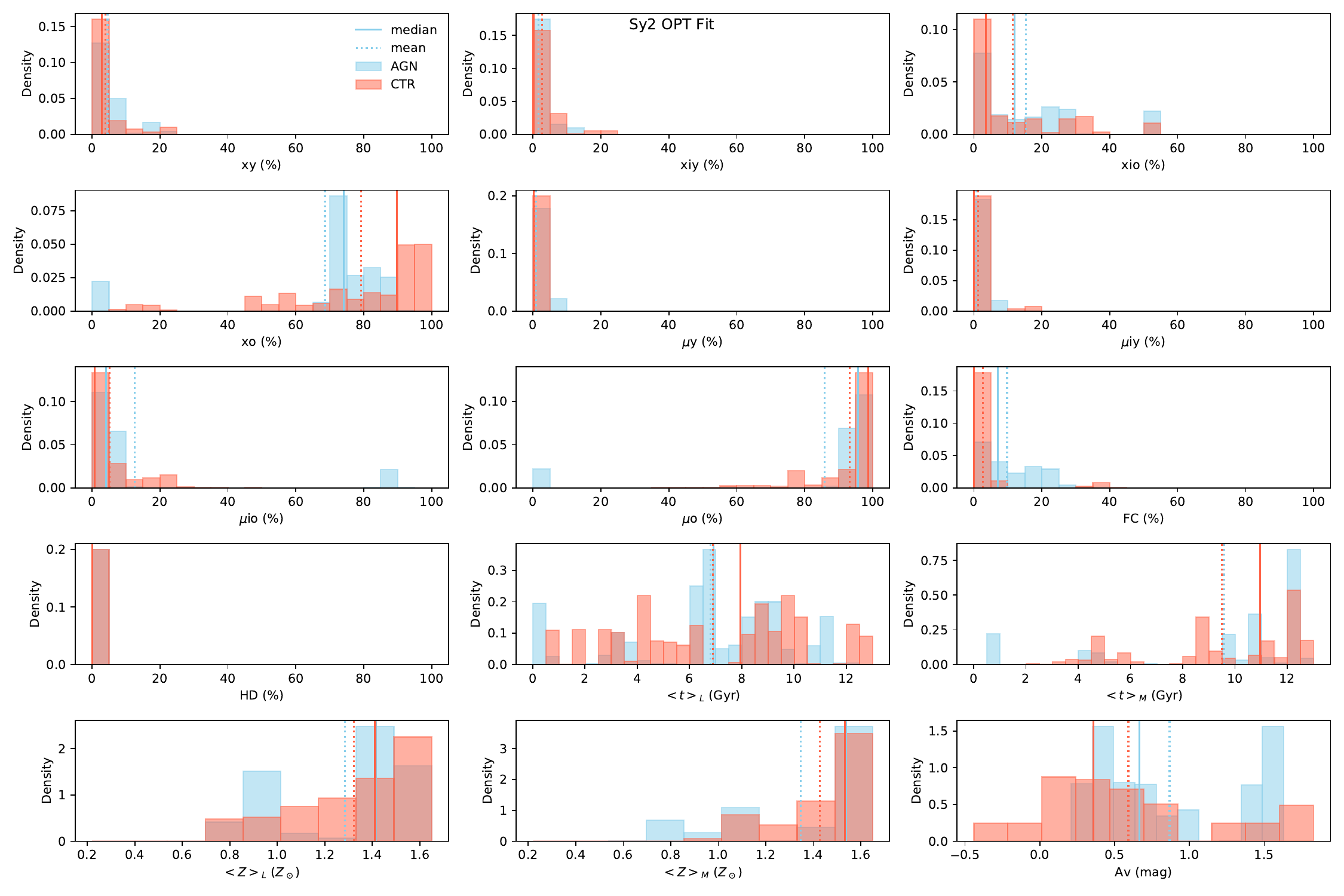}
\caption{Mean stellar population properties of the Sy~2 sample obtained after fitting UV/Optical spectral region of the 100 simulated spectra (see \S~\ref{error}). In blue AGN hosts and red control galaxies. The mean (dotted line) and the median (full line) values are shown. }
\label{fig:OPTMeanProps}
\includegraphics[width=0.9\textwidth]{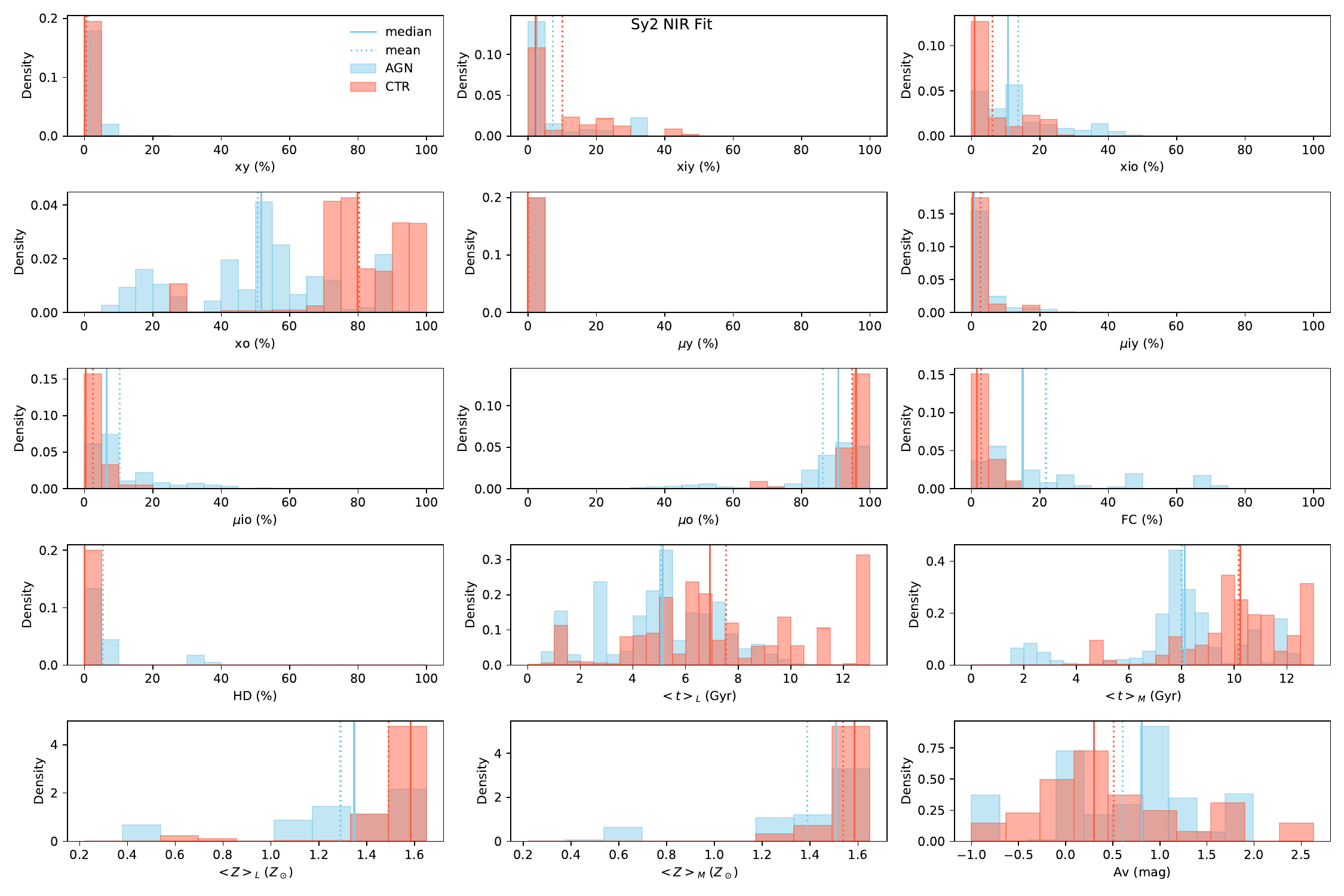}
\caption{Same as Fig.~\ref{fig:NIRMeanProps}, but for the NIR spectral region.}
\label{fig:Sy2NIRMeanProps}
\end{figure*}

The stellar populations, using the UV/Optical arms of X-shooter spectra, of part of the present sample (mostly Sy~2 sources) of LLAMA AGN have been studied by \citet{Burtscher+21}, however, using the \citet{Bruzual+03} SSP models, including ages younger than 50~Myr in the base of elements. Aiming to compare the optical and NIR stellar population results fairly, we have fitted the X-shooter UV/Optical\footnote{From 3850\AA\ to 8800\AA\ with $\lambda_{norm}$=6840\AA.} using the same approach as used for the NIR spectral range described above. The results of these fittings are shown in Fig.~\ref{fig:OPTMeanProps} and summarised in Tab.~\ref{tab:spfits_opt}. It is also worth mentioning that the results obtained by \citet{Burtscher+21} using the \citet{Bruzual+03} models are in good agreement with those found here when fitting the X-shooter data with the XSL models (see \S~\ref{stelMass} for a comparison on the derived stellar masses).

\begin{table*}
\centering
\renewcommand{\tabcolsep}{.85mm}
\caption{Properties obtained fitting the NIR spectral range.}
\label{tab:spfits}
\begin{scriptsize}
\begin{tabular}{lcccccccccccccccccc}
\noalign{\smallskip}
\hline
Source   &   xy   &  xiy   &  xio   &  xo   &  $\mu$y   &  $\mu$iy   &  $\mu$io   &  $\mu$o   &  FC    &  HD   & $<t>_L$   &  $<t>_M$  & $<Z>_L$        &  $<Z>_L$   &  $ log M_\star$ & log M$_{\star}^{Ret}$        &   Av & $\chi^2_{red}$   \\
         & (\%)   & (\%)   & (\%)   & (\%)  & (\%)      & (\%)       & (\%)       & (\%)      & (\%)    & (\%) & (Gyr)      & (Gyr)    & ($Z_{\odot}$)  & 	($Z_{\odot}$)       &  (M$_\odot$) &(M$_\odot$) & (mag) &   \\  
\noalign{\smallskip}
\multicolumn{18}{c}{Sample} \\
\noalign{\smallskip}
\hline
\noalign{\smallskip}
ESO021-G004 & 7$\pm$1   &   5$\pm$2   &   0$\pm$0   &   88$\pm$1   &   1$\pm$0   &   1$\pm$0   &   0$\pm$0   &   99$\pm$0   &   1$\pm$0   &   0$\pm$0   &   7.26$\pm$0.39   &   11.80$\pm$0.11   &   1.48$\pm$0.01   &   1.57$\pm$0.00   &   9.96$\pm$0.56   &   10.00$\pm$0.54   &   0.02$\pm$0.02   &   8$\pm$0   \\
ESO137-G034 & 0$\pm$0   &   33$\pm$0   &   10$\pm$1   &   51$\pm$0   &   0$\pm$0   &   8$\pm$0   &   6$\pm$0   &   86$\pm$0   &   5$\pm$0   &   0$\pm$0   &   2.57$\pm$0.04   &   8.08$\pm$0.08   &   1.12$\pm$0.01   &   1.47$\pm$0.00   &   9.04$\pm$0.60   &   9.05$\pm$0.57   &   -1.00$\pm$0.00   &   7$\pm$0   \\
MCG-05-14-012 & 1$\pm$3   &   0$\pm$1   &   6$\pm$8   &   32$\pm$7   &   0$\pm$1   &   0$\pm$1   &   9$\pm$12   &   91$\pm$12   &   61$\pm$4   &   0$\pm$0   &   6.25$\pm$2.66   &   7.58$\pm$2.26   &   0.84$\pm$0.43   &   0.81$\pm$0.45   &   8.40$\pm$0.48   &   8.43$\pm$0.45   &   -0.00$\pm$0.15   &   1$\pm$0   \\
MCG-05-23-016 & 0$\pm$0   &   1$\pm$2   &   10$\pm$5   &   21$\pm$4   &   0$\pm$0   &   1$\pm$2   &   14$\pm$9   &   85$\pm$9   &   68$\pm$2   &   0$\pm$0   &   4.78$\pm$1.45   &   7.02$\pm$1.23   &   1.56$\pm$0.04   &   1.57$\pm$0.02   &   8.92$\pm$0.49   &   8.90$\pm$0.46   &   0.03$\pm$0.08   &   1$\pm$0   \\
MCG-06-30-015 & 0$\pm$0   &   7$\pm$2   &   8$\pm$2   &   31$\pm$5   &   0$\pm$0   &   6$\pm$2   &   8$\pm$3   &   87$\pm$4   &   50$\pm$2   &   5$\pm$0   &   3.31$\pm$0.53   &   5.76$\pm$0.58   &   1.28$\pm$0.04   &   1.41$\pm$0.03   &   8.81$\pm$0.48   &   8.78$\pm$0.46   &   0.83$\pm$0.04   &   11$\pm$0   \\
NGC1365 & 0$\pm$0   &   0$\pm$0   &   0$\pm$0   &   21$\pm$1   &   0$\pm$0   &   0$\pm$0   &   0$\pm$0   &   100$\pm$0   &   76$\pm$1   &   2$\pm$0   &   9.95$\pm$0.66   &   10.40$\pm$0.30   &   1.25$\pm$0.04   &   1.26$\pm$0.04   &   8.65$\pm$0.52   &   8.68$\pm$0.49   &   1.29$\pm$0.06   &   11$\pm$0   \\
NGC2110 & 0$\pm$0   &   2$\pm$1   &   10$\pm$2   &   41$\pm$1   &   0$\pm$0   &   1$\pm$1   &   7$\pm$1   &   92$\pm$1   &   47$\pm$1   &   0$\pm$0   &   5.20$\pm$0.26   &   7.70$\pm$0.29   &   1.58$\pm$0.00   &   1.58$\pm$0.00   &   9.42$\pm$0.52   &   9.42$\pm$0.50   &   0.83$\pm$0.03   &   2$\pm$0   \\
NGC2992 & 2$\pm$2   &   2$\pm$1   &   0$\pm$0   &   69$\pm$2   &   0$\pm$0   &   0$\pm$0   &   0$\pm$0   &   99$\pm$0   &   19$\pm$1   &   8$\pm$1   &   8.33$\pm$0.91   &   10.70$\pm$0.24   &   1.26$\pm$0.03   &   1.30$\pm$0.02   &   9.73$\pm$0.51   &   9.76$\pm$0.49   &   1.01$\pm$0.09   &   16$\pm$0   \\
NGC3081 & 0$\pm$0   &   2$\pm$3   &   16$\pm$4   &   62$\pm$11   &   0$\pm$0   &   1$\pm$1   &   8$\pm$3   &   91$\pm$3   &   18$\pm$7   &   1$\pm$1   &   5.20$\pm$0.95   &   7.71$\pm$0.56   &   1.58$\pm$0.00   &   1.58$\pm$0.00   &   9.20$\pm$0.48   &   9.20$\pm$0.46   &   0.53$\pm$0.10   &   8$\pm$0   \\
NGC3783 & 3$\pm$2   &   10$\pm$3   &   8$\pm$2   &   25$\pm$5   &   1$\pm$1   &   8$\pm$4   &   9$\pm$5   &   82$\pm$8   &   47$\pm$3   &   7$\pm$0   &   2.29$\pm$0.52   &   5.72$\pm$0.76   &   0.82$\pm$0.06   &   0.99$\pm$0.05   &   9.17$\pm$0.48   &   9.15$\pm$0.46   &   -0.28$\pm$0.06   &   18$\pm$0   \\
NGC4235 & 2$\pm$2   &   24$\pm$7   &   8$\pm$6   &   26$\pm$6   &   1$\pm$1   &   15$\pm$6   &   10$\pm$8   &   74$\pm$10   &   34$\pm$3   &   6$\pm$1   &   1.38$\pm$0.51   &   4.28$\pm$1.10   &   1.31$\pm$0.10   &   1.46$\pm$0.06   &   8.98$\pm$0.48   &   8.93$\pm$0.45   &   1.78$\pm$0.17   &   1$\pm$0   \\
NGC4388 & 0$\pm$0   &   6$\pm$2   &   0$\pm$0   &   74$\pm$4   &   0$\pm$0   &   2$\pm$1   &   0$\pm$0   &   98$\pm$1   &   16$\pm$2   &   5$\pm$0   &   6.60$\pm$0.50   &   8.29$\pm$0.29   &   1.58$\pm$0.00   &   1.58$\pm$0.00   &   9.25$\pm$0.50   &   9.25$\pm$0.48   &   2.45$\pm$0.03   &   8$\pm$0   \\
NGC4593 & 7$\pm$2   &   1$\pm$2   &   0$\pm$0   &   28$\pm$5   &   3$\pm$2   &   0$\pm$1   &   0$\pm$0   &   96$\pm$3   &   40$\pm$1   &   24$\pm$1   &   3.36$\pm$1.22   &   8.27$\pm$0.89   &   0.62$\pm$0.05   &   0.74$\pm$0.07   &   8.74$\pm$0.48   &   8.77$\pm$0.46   &   0.59$\pm$0.15   &   12$\pm$0   \\
NGC5128 & 0$\pm$0   &   0$\pm$0   &   0$\pm$0   &   100$\pm$0   &   0$\pm$0   &   0$\pm$0   &   0$\pm$0   &   100$\pm$0   &   0$\pm$0   &   0$\pm$0   &   8.40$\pm$0.19   &   9.47$\pm$0.14   &   1.58$\pm$0.00   &   1.58$\pm$0.00   &   8.26$\pm$0.56   &   8.27$\pm$0.53   &   3.18$\pm$0.00   &   17$\pm$0   \\
NGC5506 & 8$\pm$4   &   1$\pm$2   &   14$\pm$6   &   24$\pm$9   &   5$\pm$4   &   1$\pm$3   &   27$\pm$16   &   66$\pm$19   &   30$\pm$3   &   23$\pm$2   &   1.53$\pm$0.73   &   2.75$\pm$0.95   &   1.04$\pm$0.14   &   1.01$\pm$0.14   &   9.37$\pm$0.46   &   9.27$\pm$0.44   &   4.97$\pm$0.15   &   1$\pm$0   \\
NGC5728 & 0$\pm$0   &   0$\pm$3   &   38$\pm$5   &   55$\pm$3   &   0$\pm$0   &   0$\pm$1   &   18$\pm$3   &   82$\pm$2   &   7$\pm$2   &   0$\pm$0   &   4.70$\pm$0.40   &   7.80$\pm$0.39   &   0.48$\pm$0.08   &   0.59$\pm$0.10   &   9.21$\pm$0.51   &   9.24$\pm$0.48   &   1.31$\pm$0.16   &   25$\pm$6   \\
NGC6814 & 0$\pm$1   &   23$\pm$3   &   0$\pm$0   &   47$\pm$3   &   0$\pm$0   &   7$\pm$1   &   0$\pm$0   &   93$\pm$1   &   26$\pm$1   &   4$\pm$1   &   2.94$\pm$0.43   &   8.20$\pm$0.33   &   1.21$\pm$0.03   &   1.37$\pm$0.02   &   8.87$\pm$0.52   &   8.88$\pm$0.49   &   0.85$\pm$0.07   &   10$\pm$0   \\
NGC7172 & 1$\pm$2   &   18$\pm$5   &   27$\pm$5   &   15$\pm$4   &   0$\pm$1   &   17$\pm$5   &   35$\pm$8   &   48$\pm$9   &   5$\pm$2   &   34$\pm$1   &   1.25$\pm$0.24   &   2.34$\pm$0.56   &   1.22$\pm$0.07   &   1.31$\pm$0.06   &   8.77$\pm$0.49   &   8.65$\pm$0.46   &   0.87$\pm$0.12   &   14$\pm$0   \\
NGC7213 & 0$\pm$0   &   13$\pm$6   &   10$\pm$6   &   38$\pm$7   &   0$\pm$0   &   7$\pm$3   &   7$\pm$10   &   86$\pm$9   &   39$\pm$4   &   0$\pm$0   &   3.75$\pm$0.86   &   7.29$\pm$0.92   &   1.43$\pm$0.13   &   1.53$\pm$0.13   &   9.17$\pm$0.49   &   9.16$\pm$0.46   &   0.36$\pm$0.08   &   1$\pm$0   \\
NGC7582 & 0$\pm$0   &   2$\pm$1   &   12$\pm$2   &   54$\pm$4   &   0$\pm$0   &   1$\pm$0   &   6$\pm$1   &   93$\pm$1   &   27$\pm$2   &   6$\pm$0   &   6.34$\pm$0.46   &   8.84$\pm$0.27   &   1.33$\pm$0.05   &   1.50$\pm$0.02   &   9.16$\pm$0.49   &   9.17$\pm$0.47   &   1.86$\pm$0.06   &   7$\pm$0   \\
\hline
 \noalign{\smallskip} 
 \multicolumn{18}{c}{Control Sample} \\ 
\noalign{\smallskip} 
 \hline 
 \noalign{\smallskip}
ESO093-G003 & 0$\pm$0   &   18$\pm$2   &   3$\pm$1   &   75$\pm$1   &   0$\pm$0   &   3$\pm$0   &   1$\pm$0   &   96$\pm$0   &   4$\pm$0   &   0$\pm$0   &   5.02$\pm$0.37   &   10.30$\pm$0.20   &   1.54$\pm$0.02   &   1.58$\pm$0.00   &   9.03$\pm$0.55   &   9.06$\pm$0.53   &   -0.25$\pm$0.02   &   5$\pm$0   \\
ESO208-G021 & 0$\pm$0   &   0$\pm$0   &   0$\pm$0   &   100$\pm$0   &   0$\pm$0   &   0$\pm$0   &   0$\pm$0   &   100$\pm$0   &   0$\pm$0   &   0$\pm$0   &   9.21$\pm$0.90   &   9.46$\pm$0.67   &   1.58$\pm$0.00   &   1.58$\pm$0.00   &   9.29$\pm$0.50   &   9.30$\pm$0.48   &   0.35$\pm$0.01   &   1$\pm$0   \\
IC4653 & 0$\pm$0   &   44$\pm$1   &   23$\pm$2   &   28$\pm$1   &   0$\pm$0   &   16$\pm$0   &   15$\pm$1   &   69$\pm$1   &   4$\pm$0   &   0$\pm$0   &   1.22$\pm$0.02   &   4.82$\pm$0.11   &   0.69$\pm$0.01   &   1.20$\pm$0.01   &   8.13$\pm$0.58   &   8.10$\pm$0.54   &   -1.00$\pm$0.00   &   3$\pm$0   \\
NGC0718 & 0$\pm$0   &   5$\pm$1   &   16$\pm$1   &   72$\pm$1   &   0$\pm$0   &   1$\pm$0   &   6$\pm$0   &   92$\pm$0   &   7$\pm$0   &   0$\pm$0   &   6.92$\pm$0.11   &   10.20$\pm$0.13   &   1.58$\pm$0.00   &   1.58$\pm$0.00   &   8.91$\pm$0.59   &   8.94$\pm$0.55   &   -0.27$\pm$0.01   &   16$\pm$0   \\
NGC1079 & 9$\pm$12   &   19$\pm$17   &   2$\pm$5   &   65$\pm$16   &   1$\pm$2   &   6$\pm$7   &   1$\pm$2   &   92$\pm$7   &   5$\pm$3   &   0$\pm$0   &   4.29$\pm$2.78   &   8.70$\pm$2.23   &   1.39$\pm$0.19   &   1.50$\pm$0.14   &   8.75$\pm$0.48   &   8.77$\pm$0.45   &   1.33$\pm$0.24   &   1$\pm$0   \\
NGC1315 & 0$\pm$0   &   0$\pm$0   &   12$\pm$2   &   88$\pm$2   &   0$\pm$0   &   0$\pm$0   &   4$\pm$1   &   96$\pm$1   &   0$\pm$0   &   0$\pm$0   &   9.20$\pm$0.46   &   11.50$\pm$0.20   &   1.58$\pm$0.02   &   1.58$\pm$0.02   &   8.39$\pm$0.56   &   8.43$\pm$0.53   &   -0.40$\pm$0.02   &   1$\pm$0   \\
NGC1947 & 0$\pm$2   &   7$\pm$6   &   3$\pm$4   &   90$\pm$5   &   0$\pm$0   &   2$\pm$2   &   1$\pm$1   &   97$\pm$1   &   0$\pm$0   &   0$\pm$0   &   6.51$\pm$0.91   &   8.29$\pm$1.14   &   1.57$\pm$0.03   &   1.58$\pm$0.01   &   8.84$\pm$0.51   &   8.84$\pm$0.48   &   0.90$\pm$0.05   &   1$\pm$0   \\
NGC2775 & 0$\pm$0   &   0$\pm$0   &   17$\pm$1   &   83$\pm$1   &   0$\pm$0   &   0$\pm$0   &   6$\pm$0   &   94$\pm$0   &   0$\pm$0   &   0$\pm$0   &   6.39$\pm$0.09   &   8.01$\pm$0.12   &   1.58$\pm$0.00   &   1.58$\pm$0.00   &   9.43$\pm$0.56   &   9.43$\pm$0.53   &   0.24$\pm$0.01   &   10$\pm$0   \\
NGC3175 & 0$\pm$0   &   14$\pm$0   &   0$\pm$0   &   75$\pm$0   &   0$\pm$0   &   4$\pm$0   &   0$\pm$0   &   96$\pm$0   &   10$\pm$0   &   0$\pm$0   &   7.70$\pm$0.12   &   11.00$\pm$0.10   &   1.57$\pm$0.00   &   1.58$\pm$0.00   &   8.32$\pm$0.61   &   8.36$\pm$0.57   &   1.63$\pm$0.01   &   7$\pm$0   \\
NGC3351 & 0$\pm$0   &   0$\pm$0   &   8$\pm$1   &   90$\pm$1   &   0$\pm$0   &   0$\pm$0   &   3$\pm$0   &   97$\pm$0   &   1$\pm$0   &   0$\pm$0   &   10.10$\pm$0.14   &   11.60$\pm$0.13   &   1.58$\pm$0.00   &   1.58$\pm$0.00   &   8.48$\pm$0.60   &   8.51$\pm$0.56   &   0.81$\pm$0.01   &   6$\pm$0   \\
NGC3717 & 0$\pm$0   &   25$\pm$1   &   0$\pm$0   &   72$\pm$1   &   0$\pm$0   &   4$\pm$0   &   0$\pm$0   &   96$\pm$0   &   1$\pm$1   &   2$\pm$0   &   3.99$\pm$0.14   &   10.20$\pm$0.15   &   1.58$\pm$0.00   &   1.58$\pm$0.00   &   9.25$\pm$0.55   &   9.28$\pm$0.52   &   2.58$\pm$0.02   &   6$\pm$0   \\
NGC3749 & 0$\pm$0   &   0$\pm$0   &   0$\pm$0   &   92$\pm$0   &   0$\pm$0   &   0$\pm$0   &   0$\pm$0   &   100$\pm$0   &   8$\pm$0   &   0$\pm$0   &   12.60$\pm$0.04   &   12.60$\pm$0.04   &   1.58$\pm$0.00   &   1.58$\pm$0.00   &   9.92$\pm$0.61   &   9.96$\pm$0.58   &   1.85$\pm$0.00   &   16$\pm$0   \\
NGC4224 & 0$\pm$0   &   23$\pm$2   &   0$\pm$0   &   77$\pm$2   &   0$\pm$0   &   3$\pm$0   &   0$\pm$0   &   97$\pm$0   &   0$\pm$0   &   0$\pm$0   &   4.69$\pm$0.47   &   10.80$\pm$0.22   &   1.46$\pm$0.02   &   1.57$\pm$0.00   &   9.31$\pm$0.54   &   9.34$\pm$0.51   &   -0.01$\pm$0.03   &   2$\pm$0   \\
NGC4254 & 0$\pm$0   &   25$\pm$1   &   0$\pm$0   &   74$\pm$0   &   0$\pm$0   &   6$\pm$0   &   0$\pm$0   &   94$\pm$0   &   2$\pm$0   &   0$\pm$0   &   5.40$\pm$0.10   &   9.89$\pm$0.13   &   1.51$\pm$0.01   &   1.57$\pm$0.00   &   8.39$\pm$0.60   &   8.41$\pm$0.56   &   0.50$\pm$0.01   &   5$\pm$0   \\
NGC4260 & 0$\pm$0   &   0$\pm$0   &   21$\pm$1   &   79$\pm$1   &   0$\pm$0   &   0$\pm$0   &   8$\pm$0   &   92$\pm$0   &   0$\pm$0   &   0$\pm$0   &   6.94$\pm$0.10   &   9.03$\pm$0.20   &   1.53$\pm$0.01   &   1.53$\pm$0.01   &   8.82$\pm$0.57   &   8.84$\pm$0.53   &   0.28$\pm$0.01   &   5$\pm$0   \\
NGC5037 & 0$\pm$0   &   0$\pm$0   &   0$\pm$0   &   94$\pm$0   &   0$\pm$0   &   0$\pm$0   &   0$\pm$0   &   100$\pm$0   &   6$\pm$0   &   0$\pm$0   &   12.60$\pm$0.03   &   12.60$\pm$0.00   &   1.42$\pm$0.02   &   1.42$\pm$0.02   &   9.11$\pm$0.59   &   9.15$\pm$0.58   &   0.65$\pm$0.01   &   3$\pm$0   \\
NGC5845 & 0$\pm$0   &   0$\pm$0   &   4$\pm$0   &   96$\pm$0   &   0$\pm$0   &   0$\pm$0   &   1$\pm$0   &   99$\pm$0   &   0$\pm$0   &   0$\pm$0   &   11.40$\pm$0.14   &   12.30$\pm$0.11   &   1.42$\pm$0.02   &   1.44$\pm$0.02   &   9.43$\pm$0.59   &   9.47$\pm$0.56   &   0.30$\pm$0.01   &   39$\pm$0   \\
NGC5921 & 0$\pm$0   &   12$\pm$1   &   5$\pm$1   &   79$\pm$1   &   0$\pm$0   &   3$\pm$0   &   2$\pm$0   &   95$\pm$0   &   4$\pm$0   &   0$\pm$0   &   6.45$\pm$0.07   &   9.54$\pm$0.17   &   1.58$\pm$0.00   &   1.58$\pm$0.00   &   9.94$\pm$0.56   &   9.95$\pm$0.52   &   0.04$\pm$0.01   &   18$\pm$0   \\
NGC7727 & 0$\pm$0   &   0$\pm$0   &   0$\pm$0   &   97$\pm$0   &   0$\pm$0   &   0$\pm$0   &   0$\pm$0   &   100$\pm$0   &   3$\pm$0   &   0$\pm$0   &   12.60$\pm$0.00   &   12.60$\pm$0.00   &   1.58$\pm$0.00   &   1.58$\pm$0.00   &   9.49$\pm$0.69   &   9.54$\pm$0.63   &   0.19$\pm$0.00   &   14$\pm$0   \\
\hline
 \end{tabular} 
 \end{scriptsize}
 \end{table*}
\begin{table*}
\centering
\renewcommand{\tabcolsep}{.85mm}
\caption{Properties obtained fitting the optical spectral range for the galaxies studied in \citet{Burtscher+21}.}
\label{tab:spfits_opt}
\begin{scriptsize}
\begin{tabular}{lcccccccccccccccccc}
\noalign{\smallskip}
\hline
Source   &   xy   &  xiy   &  xio   &  xo   &  $\mu$y   &  $\mu$iy   &  $\mu$io   &  $\mu$o   &  FC    &  HD   & $<t>_L$   &  $<t>_M$  & $<Z>_L$        &  $<Z>_L$   &  $ log M_\star$ & log M$_{\star}^{Ret}$        &   Av & $\chi^2_{red}$   \\
         & (\%)   & (\%)   & (\%)   & (\%)  & (\%)      & (\%)       & (\%)       & (\%)      & (\%)    & (\%) & (Gyr)      & (Gyr)    & ($Z_{\odot}$)  & 	($Z_{\odot}$)       &  (M$_\odot$) &(M$_\odot$) & (mag) &   \\  
\noalign{\smallskip}
\multicolumn{18}{c}{Sample} \\
\noalign{\smallskip}
\hline
\noalign{\smallskip}
NGC2110 & 6$\pm$1   &   0$\pm$0   &   0$\pm$0   &   71$\pm$1   &   0$\pm$0   &   0$\pm$0   &   0$\pm$0   &   100$\pm$0   &   24$\pm$2   &   0$\pm$0   &   8.38$\pm$0.38   &   12.30$\pm$0.04   &   0.02$\pm$0.00   &   0.02$\pm$0.00   &   8.97$\pm$0.53   &   9.01$\pm$0.50   &   0.56$\pm$0.01   &   1$\pm$0   \\
NGC2992 & 0$\pm$0   &   0$\pm$0   &   5$\pm$1   &   89$\pm$0   &   0$\pm$0   &   0$\pm$0   &   1$\pm$0   &   99$\pm$0   &   6$\pm$1   &   0$\pm$0   &   11.20$\pm$0.23   &   12.30$\pm$0.04   &   0.01$\pm$0.00   &   0.02$\pm$0.00   &   9.25$\pm$0.58   &   9.31$\pm$0.56   &   1.61$\pm$0.01   &   2$\pm$0   \\
MCG-05-23-016 & 9$\pm$1   &   0$\pm$0   &   0$\pm$0   &   74$\pm$1   &   0$\pm$0   &   0$\pm$0   &   0$\pm$0   &   100$\pm$0   &   16$\pm$1   &   1$\pm$0   &   6.81$\pm$0.23   &   12.10$\pm$0.16   &   0.02$\pm$0.00   &   0.02$\pm$0.00   &   8.71$\pm$0.56   &   8.75$\pm$0.53   &   0.66$\pm$0.02   &   2$\pm$0   \\
NGC3081 & 5$\pm$0   &   0$\pm$0   &   0$\pm$0   &   81$\pm$0   &   0$\pm$0   &   0$\pm$0   &   0$\pm$0   &   100$\pm$0   &   14$\pm$1   &   0$\pm$0   &   9.25$\pm$0.23   &   12.30$\pm$0.20   &   0.02$\pm$0.00   &   0.02$\pm$0.00   &   8.91$\pm$0.58   &   8.95$\pm$0.54   &   0.48$\pm$0.01   &   2$\pm$0   \\
ESO021-G004 & 3$\pm$3   &   2$\pm$3   &   13$\pm$6   &   81$\pm$4   &   0$\pm$0   &   0$\pm$0   &   5$\pm$3   &   94$\pm$3   &   1$\pm$2   &   0$\pm$0   &   3.42$\pm$0.44   &   4.67$\pm$0.56   &   0.02$\pm$0.00   &   0.02$\pm$0.00   &   8.94$\pm$0.50   &   8.89$\pm$0.47   &   0.92$\pm$0.04   &   1$\pm$0   \\
NGC5728 & 0$\pm$0   &   0$\pm$0   &   28$\pm$1   &   72$\pm$1   &   0$\pm$0   &   0$\pm$0   &   7$\pm$0   &   93$\pm$0   &   0$\pm$0   &   0$\pm$0   &   6.40$\pm$0.12   &   9.84$\pm$0.17   &   0.02$\pm$0.00   &   0.02$\pm$0.00   &   9.16$\pm$0.55   &   9.18$\pm$0.52   &   0.39$\pm$0.01   &   2$\pm$0   \\
ESO137-G034 & 0$\pm$0   &   0$\pm$0   &   17$\pm$3   &   77$\pm$2   &   0$\pm$0   &   0$\pm$0   &   5$\pm$1   &   95$\pm$1   &   7$\pm$2   &   0$\pm$0   &   8.61$\pm$0.49   &   10.90$\pm$0.28   &   0.01$\pm$0.00   &   0.02$\pm$0.00   &   8.99$\pm$0.55   &   9.03$\pm$0.52   &   0.26$\pm$0.02   &   2$\pm$0   \\
NGC7172 & 0$\pm$0   &   4$\pm$1   &   23$\pm$2   &   73$\pm$1   &   0$\pm$0   &   0$\pm$0   &   6$\pm$0   &   94$\pm$0   &   0$\pm$0   &   0$\pm$0   &   6.69$\pm$0.14   &   11.00$\pm$0.05   &   0.02$\pm$0.00   &   0.02$\pm$0.00   &   9.02$\pm$0.56   &   9.05$\pm$0.54   &   1.53$\pm$0.01   &   2$\pm$0   \\
NGC7582 & 18$\pm$2   &   9$\pm$3   &   52$\pm$1   &   0$\pm$0   &   6$\pm$1   &   6$\pm$2   &   88$\pm$1   &   0$\pm$0   &   20$\pm$1   &   0$\pm$0   &   0.49$\pm$0.02   &   0.94$\pm$0.02   &   0.01$\pm$0.00   &   0.01$\pm$0.00   &   8.16$\pm$0.56   &   7.92$\pm$0.52   &   1.39$\pm$0.02   &   3$\pm$0   \\
\hline
 \noalign{\smallskip} 
 \multicolumn{18}{c}{Control Sample} \\ 
\noalign{\smallskip} 
 \hline 
 \noalign{\smallskip}
ESO093-G003 & 4$\pm$1   &   0$\pm$1   &   0$\pm$0   &   96$\pm$1   &   0$\pm$0   &   0$\pm$0   &   0$\pm$0   &   100$\pm$0   &   0$\pm$0   &   0$\pm$0   &   4.90$\pm$0.17   &   8.55$\pm$0.29   &   0.01$\pm$0.00   &   0.02$\pm$0.00   &   8.72$\pm$0.54   &   8.74$\pm$0.51   &   0.76$\pm$0.01   &   2$\pm$0   \\
ESO208-G021 & 5$\pm$1   &   1$\pm$1   &   0$\pm$0   &   93$\pm$0   &   0$\pm$0   &   0$\pm$0   &   0$\pm$0   &   100$\pm$0   &   0$\pm$0   &   0$\pm$0   &   8.95$\pm$0.16   &   12.30$\pm$0.00   &   0.02$\pm$0.00   &   0.02$\pm$0.00   &   8.35$\pm$0.59   &   8.39$\pm$0.56   &   0.68$\pm$0.01   &   2$\pm$0   \\
IC4653 & 13$\pm$3   &   20$\pm$2   &   16$\pm$6   &   14$\pm$3   &   3$\pm$1   &   16$\pm$2   &   20$\pm$9   &   62$\pm$9   &   37$\pm$2   &   0$\pm$0   &   0.78$\pm$0.09   &   3.81$\pm$0.63   &   0.01$\pm$0.00   &   0.02$\pm$0.00   &   7.11$\pm$0.51   &   7.05$\pm$0.48   &   -0.38$\pm$0.03   &   2$\pm$0   \\
NGC0718 & 0$\pm$0   &   0$\pm$0   &   18$\pm$1   &   82$\pm$1   &   0$\pm$0   &   0$\pm$0   &   10$\pm$0   &   90$\pm$0   &   0$\pm$0   &   0$\pm$0   &   4.36$\pm$0.03   &   4.79$\pm$0.03   &   0.02$\pm$0.00   &   0.02$\pm$0.00   &   8.85$\pm$0.64   &   8.82$\pm$0.58   &   0.53$\pm$0.00   &   1$\pm$0   \\
NGC1079 & 4$\pm$0   &   0$\pm$0   &   4$\pm$1   &   90$\pm$1   &   0$\pm$0   &   0$\pm$0   &   1$\pm$0   &   99$\pm$0   &   2$\pm$1   &   0$\pm$0   &   8.40$\pm$0.22   &   11.40$\pm$0.14   &   0.02$\pm$0.00   &   0.02$\pm$0.00   &   8.54$\pm$0.56   &   8.57$\pm$0.53   &   0.25$\pm$0.02   &   1$\pm$0   \\
NGC1315 & 4$\pm$0   &   0$\pm$0   &   0$\pm$1   &   95$\pm$0   &   0$\pm$0   &   0$\pm$0   &   0$\pm$0   &   100$\pm$0   &   0$\pm$0   &   0$\pm$0   &   9.42$\pm$0.11   &   12.10$\pm$0.16   &   0.02$\pm$0.00   &   0.02$\pm$0.00   &   8.32$\pm$0.58   &   8.38$\pm$0.55   &   0.20$\pm$0.00   &   1$\pm$0   \\
NGC1947 & 5$\pm$1   &   1$\pm$1   &   6$\pm$2   &   89$\pm$1   &   0$\pm$0   &   0$\pm$0   &   1$\pm$0   &   98$\pm$0   &   0$\pm$0   &   0$\pm$0   &   8.44$\pm$0.14   &   12.10$\pm$0.12   &   0.02$\pm$0.00   &   0.02$\pm$0.00   &   8.32$\pm$0.58   &   8.36$\pm$0.55   &   1.43$\pm$0.01   &   2$\pm$0   \\
NGC2775 & 0$\pm$0   &   0$\pm$0   &   0$\pm$0   &   100$\pm$0   &   0$\pm$0   &   0$\pm$0   &   0$\pm$0   &   100$\pm$0   &   0$\pm$0   &   0$\pm$0   &   12.50$\pm$0.13   &   12.60$\pm$0.00   &   0.02$\pm$0.00   &   0.02$\pm$0.00   &   8.42$\pm$0.67   &   8.47$\pm$0.63   &   0.19$\pm$0.00   &   2$\pm$0   \\
NGC3175 & 22$\pm$1   &   8$\pm$2   &   0$\pm$0   &   70$\pm$1   &   2$\pm$0   &   1$\pm$0   &   0$\pm$0   &   97$\pm$0   &   0$\pm$0   &   0$\pm$0   &   1.83$\pm$0.05   &   8.55$\pm$0.16   &   0.02$\pm$0.00   &   0.02$\pm$0.00   &   7.99$\pm$0.58   &   8.01$\pm$0.54   &   1.21$\pm$0.01   &   1$\pm$0   \\
NGC3351 & 3$\pm$1   &   1$\pm$1   &   10$\pm$2   &   79$\pm$1   &   0$\pm$0   &   0$\pm$0   &   3$\pm$0   &   97$\pm$0   &   7$\pm$1   &   0$\pm$0   &   5.72$\pm$0.30   &   8.76$\pm$0.36   &   0.02$\pm$0.00   &   0.02$\pm$0.00   &   7.28$\pm$0.55   &   7.29$\pm$0.52   &   0.52$\pm$0.01   &   1$\pm$0   \\
NGC3717 & 1$\pm$1   &   7$\pm$2   &   32$\pm$2   &   59$\pm$1   &   0$\pm$0   &   1$\pm$0   &   10$\pm$1   &   89$\pm$1   &   0$\pm$0   &   0$\pm$0   &   4.34$\pm$0.11   &   9.39$\pm$0.16   &   0.02$\pm$0.00   &   0.02$\pm$0.00   &   8.59$\pm$0.56   &   8.61$\pm$0.53   &   1.65$\pm$0.01   &   1$\pm$0   \\
NGC3749 & 0$\pm$0   &   0$\pm$0   &   27$\pm$1   &   73$\pm$1   &   0$\pm$0   &   0$\pm$0   &   8$\pm$0   &   92$\pm$0   &   0$\pm$0   &   0$\pm$0   &   6.09$\pm$0.11   &   8.75$\pm$0.10   &   0.02$\pm$0.00   &   0.02$\pm$0.00   &   9.20$\pm$0.57   &   9.21$\pm$0.54   &   1.82$\pm$0.01   &   2$\pm$0   \\
NGC4224 & 0$\pm$0   &   1$\pm$0   &   4$\pm$1   &   92$\pm$1   &   0$\pm$0   &   0$\pm$0   &   1$\pm$0   &   99$\pm$0   &   2$\pm$0   &   0$\pm$0   &   9.91$\pm$0.17   &   11.10$\pm$0.12   &   0.02$\pm$0.00   &   0.02$\pm$0.00   &   9.18$\pm$0.59   &   9.21$\pm$0.55   &   0.35$\pm$0.00   &   1$\pm$0   \\
NGC4254 & 0$\pm$0   &   0$\pm$0   &   53$\pm$1   &   47$\pm$1   &   0$\pm$0   &   0$\pm$0   &   22$\pm$1   &   78$\pm$1   &   0$\pm$0   &   0$\pm$0   &   3.43$\pm$0.07   &   5.81$\pm$0.16   &   0.02$\pm$0.00   &   0.02$\pm$0.00   &   8.45$\pm$0.56   &   8.41$\pm$0.52   &   0.28$\pm$0.01   &   1$\pm$0   \\
NGC5037 & 3$\pm$0   &   0$\pm$0   &   0$\pm$0   &   96$\pm$1   &   0$\pm$0   &   0$\pm$0   &   0$\pm$0   &   100$\pm$0   &   1$\pm$1   &   0$\pm$0   &   9.97$\pm$0.34   &   11.90$\pm$0.26   &   0.02$\pm$0.00   &   0.02$\pm$0.00   &   8.54$\pm$0.56   &   8.59$\pm$0.53   &   0.89$\pm$0.01   &   2$\pm$0   \\
NGC5845 & 2$\pm$0   &   1$\pm$0   &   2$\pm$0   &   95$\pm$0   &   0$\pm$0   &   0$\pm$0   &   0$\pm$0   &   100$\pm$0   &   0$\pm$0   &   1$\pm$0   &   10.30$\pm$0.12   &   12.10$\pm$0.15   &   0.02$\pm$0.00   &   0.02$\pm$0.00   &   8.87$\pm$0.59   &   8.90$\pm$0.55   &   0.03$\pm$0.00   &   2$\pm$0   \\
NGC5921 & 5$\pm$0   &   8$\pm$1   &   32$\pm$3   &   56$\pm$2   &   0$\pm$0   &   2$\pm$0   &   18$\pm$2   &   79$\pm$2   &   0$\pm$0   &   0$\pm$0   &   2.59$\pm$0.05   &   4.82$\pm$0.15   &   0.02$\pm$0.00   &   0.02$\pm$0.00   &   8.61$\pm$0.55   &   8.57$\pm$0.52   &   0.29$\pm$0.00   &   1$\pm$0   \\
NGC7727 & 0$\pm$0   &   0$\pm$0   &   0$\pm$0   &   100$\pm$0   &   0$\pm$0   &   0$\pm$0   &   0$\pm$0   &   100$\pm$0   &   0$\pm$0   &   0$\pm$0   &   12.30$\pm$0.17   &   12.40$\pm$0.19   &   0.02$\pm$0.00   &   0.02$\pm$0.00   &   8.87$\pm$0.58   &   8.91$\pm$0.54   &   -0.03$\pm$0.01   &   1$\pm$0   \\
\hline
 \end{tabular} 
 \end{scriptsize}
 \end{table*}

\subsubsection{Optical {\it versus} NIR fits}

The results of our optical and NIR fits are in very good agreement (Fig.~\ref{fig:OPTMeanProps} and \ref{fig:Sy2NIRMeanProps}) in terms of the stellar population ages. The main difference between optical and NIR results is in the metallicities and FC components. 

The NIR fits, in general, show higher metalicities than the optical ones, especially when looking at the control galaxies. The optical results show small differences in $Z$ between controls and AGN with both samples spanning a large range of metallicities (on average AGN presents a slightly higher metallicity than controls). On the other hand, the NIR fits do show a larger difference between mean metallicities in both samples. We attribute this difference to the fact that the optical spectral range has more features sensitive to metallicity than the NIR window and/or to the higher SNR of the optical data when compared with the NIR ones. Regarding the $FC$ components in the NIR, we see a tail towards a larger fraction in the distribution of FC fractions in AGN when compared with the optical results. This may be related to the degeneracy between the $FC$ and $HD$ components \citep[see][for details]{Riffel+22}. Finally, it is worth mentioning that in the NIR we can fit also the type~1 sources.

\subsubsection{Stellar population components}
What emerges from this is that the AGN hosts have larger fractions of intermediate-age, less metallic stellar populations when compared with the control sample. Additionally, the AGN show higher reddening and, as expected only they require significant fractions of FC and HD components to fit their underlying spectrum. 

In \citet{Burtscher+21} we have shown that the LLAMA AGN show, in general, a recent cessation of their star formation, i.e. it happened at least 6\,Myr ago. We cannot examine this here because of the very young age limitations in our base (see \S~\ref{popstar}), but in the younger stellar populations bin, we do not see any significant difference in both samples, with AGN exhibiting only a slightly higher mean value for $xy$.

Our results (Fig.~\ref{fig:NIRMeanProps}) show that AGN present a lower fraction of the old stellar population when compared with controls ($\mu o$ panel). The lower $\mu o$ fractions are therefore compensated by the higher fraction of younger stellar populations\footnote{Note that in the case of the light-weighted fractions, where the difference between the AGN and controls are even more pronounced, one has to have in mind that besides the SSP components the FC and HD also contribute, however, they have no contribution in the stellar mass fraction.} (e.g. one fraction comes down, the other has to go up), thus, AGN present larger fractions of intermediate-age populations when compared with the control sources. This is directly reflected in the \maL\  mean/median values (4.7/4.5~Gyr for AGN and  7.5/6.9~Gyr for the controls) as well as in the \maM\ mean/median values  (7.5/8.0~Gyr for AGN and  10.2/10.2~Gyr for the controls).

Based on high angular resolution integral field observations, \citet{Riffel+22} found that the inner region (R$\lesssim$125 pc)  of AGN is primarily composed of intermediate-age SP (with an average age $<t>_L \lesssim 1.5$ Gyr). They also observed a correlation between the bolometric luminosity of AGN and the mean age of their stellar populations, indicating that more luminous AGN tend to have larger amounts of intermediate-age stars. This has been interpreted as suggesting a delay between the formation of new stars and the triggering/feeding of the AGN. In this sense, these intermediate-age stars contribute to the gas supply around the supermassive black hole through mass loss during stellar evolution. This gas, which has a low velocity (a few hundred km/s), combines with the existing gas which is flowing towards the central region of the host galaxy \citep[e.g.,][]{Cuadra+06,Fathi+06,Davies+07,Riffel+08a,MullerSanchez+09,Riffel+13a,Storchi-Bergmann+19,Diniz+19,Audibert+19,Audibert+21,Riffel+22, Riffel+23} and feeding the SMBH.

Furthermore, studies have detected young to intermediate-age populations in the inner regions of AGN host galaxies \citep[e.g.,][]{Oliva+95,GonzalezDelgado+01,Imanishi+04, Davies+05, Davies+07, Riffel+07, Riffel+09, Riffel+10,  Riffel+11,Riffel+15,Storchi-Bergmann+12, Mallmann+18, Diniz+17, deLorenzo-Caceres+20, Salvador-Rusinol+21, Burtscher+21, Dahmer-Hahn+22, Riffel+22, Riffel+23}. These populations are predominantly composed of short-lived stars ($t\simeq 0.2-2$ Gyr; $\rm M\simeq 2-6\,M_{\odot}$) that release significant amounts of material into the nuclear environment \citep[from $\sim$30\% - 80\% of their masses,][]{Bertolami+16}.  This material can fuel the SMBH with additional gas, thereby enhancing AGN brightness or triggering its activity or alternatively, it can cool down and form new stars \citep[e.g.,][]{Salvador-Rusinol+20, Salvador-Rusinol+21, deLorenzo-Caceres+20, Benedetti+23}.

\subsubsection{The FC and HD components}\label{sec:hd}

As can be seen in Tab.~\ref{tab:spfits} the $FC$ component is required to fit the spectral energy distribution of almost all the AGN, being the only exception NGC~5128. This component is also required in 63\% (12/19 sources) to fit the control galaxies, with a maximum contribution of 10\%.  The degeneracy of the $FC$ and a young stellar population is a well-known and common problem in the study of the stellar content of active galaxies \citep[e.g.][and references therein]{Koski+78,CidFernandes+04,CidFernandes+05,Riffel+09}. This degeneracy is due to the fact that the continuum of a reddened young starburst (t $\lesssim$ 5\,Myr) is indistinguishable from an AGN-type continuum. In the optical, broad components in polarised light of Sy~2 galaxies only show up when this component is larger than $\sim$20\% \citep{CidFernandes+95}. Taking our results into account, one could say that in the NIR the young and $FC$ components can be separated for values larger than 10\% (the maximum value reached in the control sample). Thus, for four AGN (ESO021-G004, ESO137-G034, NGC5728, and NGC~7172) the $FC$ component could not be distinguished from a young stellar population component. This is in agreement with the results 
presented in \citet{Riffel+22} who found that the $FC$ can be associated with the accretion disc emission only for sources with $FC \gtrsim$ 15\%.

The $HD$ component is necessary to fit 60\% (12/20) of the AGN, while it is only required in one of the control galaxies (NGC~3717), with a very low contribution (2\%, within the uncertainties of our fits, see Supplementary Material). This component is very common in type~1 sources and is detected in around 50\% of type~2 sources \citep[e.g.][]{Riffel+06,Riffel+09,Riffel+10,Martins+10,Gaspar+19,Riffel+22}.
For caveats on hot dust {\it versus} featureless continuum fitting see \citet{Riffel+22}.

\subsubsection{Comparison of Seyfert~1 and Seyfert~2 fits}

\begin{figure*}
\includegraphics[width=0.9\textwidth]{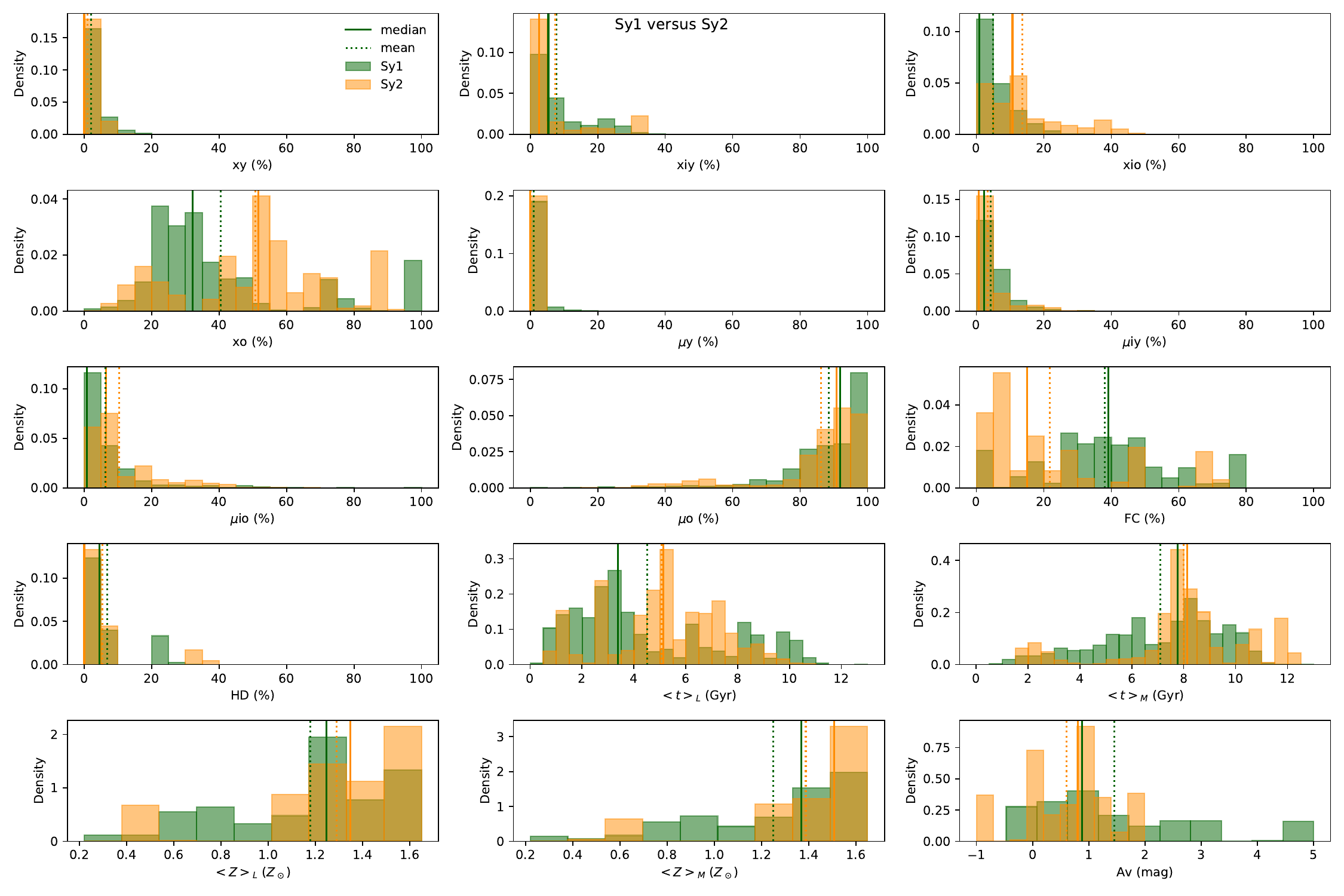}
\caption{Mean stellar population properties of the Sy~1(green) and Sy~2 (yellow) sources obtained after fitting the NIR spectral region of the 100 simulated spectra (see \S~\ref{error}).  The mean (dotted line) and the median (full line) values are shown. }
\label{fig:sy1sy2}
\end{figure*}

In Fig.~\ref{fig:sy1sy2} we compare the fitting results obtained for Sy~1 (green) and Sy~2 (yellow). In general, type~1 objects show higher fractions of $xy$ and $xiy$ than the type~2 objects, while Sy~2 sources present higher fractions of the older populations ($xio$ and $xo$). In terms of mean ages, the Sy~1 galaxies display a mean value of $<t>\sim$3\,Gyr while the type~2 objects are characterized by slightly older mean ages ($<t>\sim$5\,Gyr). In terms of mass fractions, the mean values of all population vectors are quite similar. The same trend is also seen in the \maM\ values. Both samples have similar reddening distributions with type~1 galaxies displaying a tail towards higher $A_V$ values. In terms of the stellar metalicities,  Sy~2 galaxies have slightly higher mean metallicities when compared with the Sy~1 ones. Finally, the $FC$ and $HD$ components are higher in type~1 sources, with the latter showing up only in one source (with a quite spread solution around its value, see \S~\ref{sec:hd} for more details). The results presented here suggest that LLAMA type~1 and type~2 AGN essentially host stars with the same ages, with type~1 sources showing a small bias to younger values.

\subsubsection{Stellar Masses comparison}\label{stelMass}

One of the derived quantities when fitting the stellar populations using \str\ is the stellar mass. The values we derived when fitting the NIR (optical) stellar populations are listed in Tab.~\ref{tab:spfits} (and in Tab.~\ref{tab:spfits_opt} for optical).

\begin{figure*}
\includegraphics[width=\textwidth]{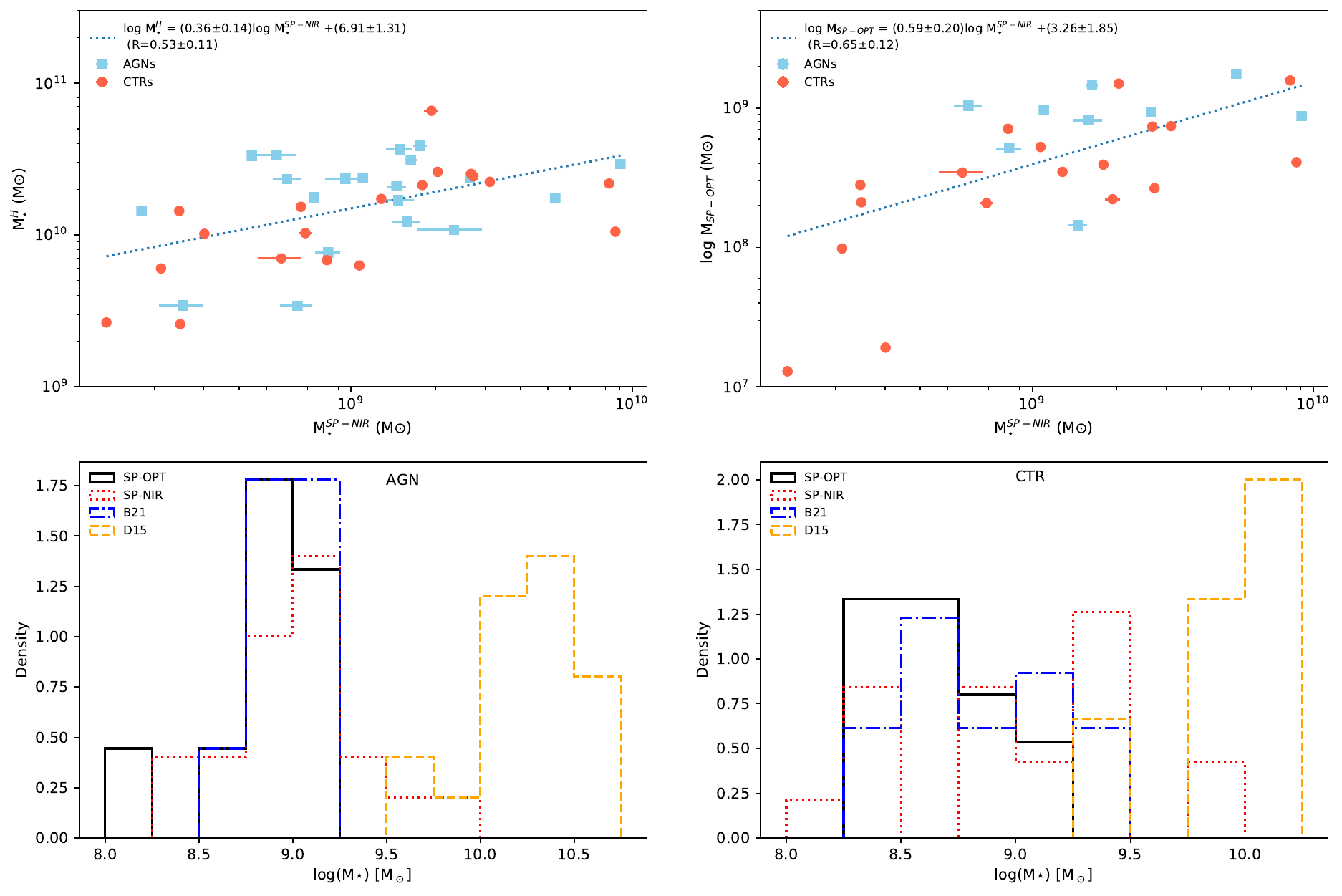}
\caption{Stellar masses comparisons. The dotted line on the top panels is a linear regression over all data points using a bootstrap scheme of 1000 realisations. In the bottom panels, the different samples are labeled as follows: SP-OPT (black) our fits using the x-shooter data from \citet[][with B21 (blue) being the fits obtained by these authors]{Burtscher+21}. SP-NIR (red) are our fits in the NIR and D15 (orange) are the masses estimated by \citet{Davies+15} for the entire galaxy. 
}
\label{fig:masses}
\end{figure*}

In Fig.~\ref{fig:masses} we compare the masses derived by \citet{Davies+15} using the $H$-band luminosity, $L_H$\footnote{Obtained using the equation log($M\star/M\odot)$~=~log($L_H/ L_\odot)$~-~0.06, which has been derived comparing multi-band photometry, to the integrated $H$-band luminosity of the galaxies studied in \citet{Koss+11}.}, with those obtained with the SP fits (top pannesl). We also compare in the bottom histograms of Fig.~\ref{fig:masses} the masses we obtained with those obtained by \citet[][B21]{Burtscher+15} for the Sy~2 of our sample  and our fits using the X-shooter data (SP-OPT) with our NIR fits (SP-NIR) and those derived by \citet[][D15]{Davies+15} for the entire galaxy. 

The median mass derived for our sample using \str\ fits of the NIR spectral range is log($M_\star/M_\odot)$=9.04 and for the optical log($M_\star/M_\odot)$=8.7. The median mass we have previously inferred from the optical \citep{Burtscher+21} is log($M_\star/M_\odot)$=8.8 and the median total mass of the galaxy inferred from the $H$-band luminosity \citep{Davies+15} is log($M_\star/M_\odot)$=10.35. These values are roughly consistent with the fact that our apertures\footnote{The aperture of the NIR data (see Tab.~\ref{tab:obs_log}) is in median 70\% of the aperture of the X-shooter data (1.8$''\times$1.8$''$).} cover 10-15\% of the mass of bulge of the LLAMA sample \citep[see][for a discussion]{Burtscher+21} which has a median bulge/total ratio of 0.25 \citep{Lin+18}. It is worth mentioning that the values we have derived with the NIR data are on average 2 times larger than those obtained with optical data (16 times lower than those derived using $L_H$, while when using the optical range the values are 35 times lower than those using the $L_H$). We attribute this to the fact that the NIR is more sensitive to the less luminous redder population, thus reflecting better the $L_H$ galaxy mass estimates.

\section{Discussion}\label{sec:discussion}
\subsection{Comparison between the stellar population and host galaxy properties}

Molecular gas serves as the essential fuel for the birth of stars. Understanding the SFH and SFR linked to each unit mass of molecular gas offers critical insights into both: the physical characteristics of the molecular phase and the surrounding conditions of star-forming regions within it.

To facilitate a comprehensive comparison of stellar population properties with the broader characteristics of host galaxies, we conducted measurements of emission line fluxes of the [\ion{O}{iii}] $\lambda$5007\AA\ and of the warm \h2\ line $\lambda$ 2.1213\mc\ on the pure emission line spectrum\footnote{With the underlying stellar population contamination subtracted.}. To do this we have employed the {\sc ifscube} code \citep{Ruschel-Dutra+21} fitting one single Gaussian to each line. 

\begin{figure*}
\includegraphics[width=\textwidth]{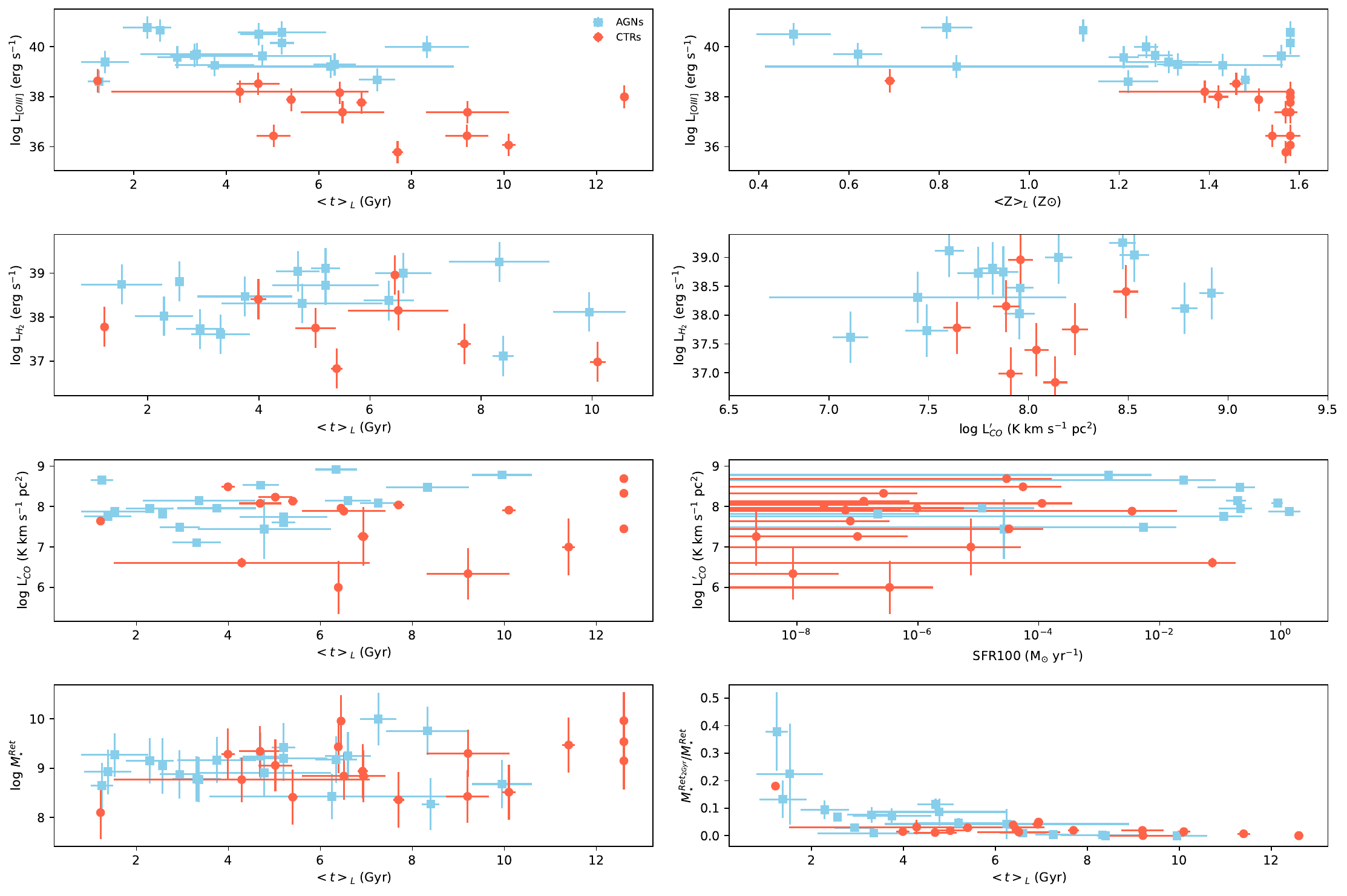}
\caption{Comparisons of stellar population properties with the host galaxy properties.
}
\label{fig:lumAges}
\end{figure*}

Additionally, following the methodology presented in \citet{Riffel+21,Riffel+23} we have computed the star formation rate over the last 100~Myr for our sources using the stellar population fits. We also collected the CO luminosities, L$'_{CO}$, from \citet{Rosario+16} for all the LLAMA galaxies (AGN and controls). The comparison of these quantities with the stellar population properties is shown in Fig.~\ref{fig:lumAges}. 

When comparing the AGN and controls distribution, it becomes clear that in terms of \oiii\ both samples show different distributions, but this is expected as discussed in \citet[][see also \citet{Riffel+23} and references therein]{Rembold+17}.  When comparing the \oiii\ luminosity with \maL\ there 
and \mzL no correlations are seen for AGN and controls. However, AGN tend to show lower ages and metallicities when compared with their controls.

The hot molecular \h2\ is not correlated with \maL. This may indicate that in AGN the excitation of the molecular gas is mostly due to AGN. In fact, this is in agreement with the fact that  \h2\ emission is mostly due to thermal excitation mechanisms associated with the AGN \citep[e.g.][]{Reunanen+02,Reunanen+03,Rodriguez-Ardila+04,Rodriguez-Ardila+05,Riffel+13,Colina+15,Riffel+21,Riffel+21b,Motter+21,Bianchin+22,Holden+23b}. 

L$'_{CO}$ shows an increasing trend with \maL, however, there is no significant difference relative to the control ones. We also found that the \h2\ luminosities are well correlated with L$'_{CO}$, with no clear separation between AGN and controls.
In general, our results show that AGN hosts tend to have higher SFR over the last 100~Myr than the control sources.

When considering the total returned mass, a difference in the amount of returned mass over the galaxy's lifetime is seen in AGN when compared with the parent control sample. However, it is worth mentioning that taking the $M_{\star}^{Ret}$ by populations younger than 2\,Gyr ($M_{\star}^{Ret (2Gyr)}$) and taking the ratio over the total returned mass ($M_{\star}^{Ret (2Gyr)}$/$M_{\star}^{Ret}$)  AGN do show a higher fraction than the control galaxies. The mean value for AGN (0.07) is $\sim$2.3 higher than the value found for the controls (0.03; see Fig.~\ref{fig:lumAges}). Additionally, when dividing by $M_{\star}$, AGN show smaller values for this ratio than the controls. This means that AGN have essentially the same $M_{\star}$ as the controls, but since they are still forming stars, they are still building their mass up, and thus this indicates that AGN are currently receiving an extra amount of gas.

To assess whether the material released by stars is sufficient to fuel the AGN, we conducted a straightforward analysis by calculating the mass accretion rate, $\dot{M}$, needed to generate the AGN luminosity. This calculation is expressed as follows:

\begin{equation}
\dot{M} = \frac{L_{\rm bol}} {\eta c^2}
\end{equation}

The bolometric luminosity, $L_{\rm bol}$, was determined based on the methodology outlined in \citet{Ichikawa+17}, given by the equation:

\begin{equation}
\log L_{\rm bol} = 0.0378 (\log L_{X:2-10})^2 - 2.00 \log L_{X:2-10} +60.5,
\label{lbol}
\end{equation}
\noindent
where the intrinsic hard X-ray luminosity from Tab.~\ref{tab:AGN} was employed. Adopting an efficiency value of $\eta = 0.1$ \citep[e.g.,][]{Laor+89,King+08}, we determined $\dot{M}$ for our galaxy sample.

Upon comparing $\dot{M}$ with the stellar mass returned (in $M_{\odot}$/yr), we observed that, on average, the returned mass from the intermediate population ($M_{\star}^{Ret(2Gyr)}$/2Gyr) is of the same order as the mass required to sustain AGN luminosity. Note however that this is the lower limit of mass that is being released since the older population ($t >$2\,Gyr) is also present in the galaxies and is also releasing mass during these last two Gyr\footnote{It is not just the intermediate age population that is releasing mass, but the older populations are also releasing mass during this time}. When considering the mean returned mass per year over the entire galaxy lifetime ($M_{\star}^{Ret}$/13\,Gyr), the returned mass is approximately 6 times the mass necessary to maintain the AGN. It is noteworthy that if this analysis were conducted using $L_X$ instead of bolometric luminosity, the resulting ratios would be approximately 20 and 120 for the mass returned over the last 2\,Gyr and the galaxy's entire lifetime, respectively.

This simple exercise clarifies that there is sufficient recycled material being released by the stars in the central region (inner 100-200\,pc) of the galaxies to sustain AGN luminosities (e.g. there is no need for extra gas to feed the AGN).

\subsection{Stellar ages and AGN properties}

We have also explored the impact/regulation that the energy injected by the AGN in the host galaxy may have on the star formation of our sources.  Therefore we have collected several properties from literature such as L$_{X}^{obs}$, L$_{X}^{int}$, N$_{H}$ which are only available for the AGN sample (listed Tab.~\ref{tab:AGN}).  We have computed the Eddington luminosity using the SMBH, $M_\bullet$, masses listed in Tab.~\ref{tab:AGN} and the following equation \citep{Rybicki+79}: 

\begin{equation}
L_{\rm Edd}=1.26\cdot 10^{46}\,\text{erg s}^{-1} \frac{M_\bullet}{10^8 M_\odot}.
\end{equation} 
Also, following \citet{Riffel+22} the Eddington ratio was obtained from $L_{Bol}$ (see eq.~\ref{lbol}) and calculated using the observed and intrinsic hard X-ray luminosities ($\lambda_{obs} = L_{Bol}^{obs}/L_{\rm Edd}$ and $\lambda_{int} = L_{Bol}^{int}/L_{\rm Edd}$).

We have compared these properties with \maL. No clear correlation is seen between these properties and the host galaxy's mean age for the stellar population, as shown by their Pearson rank correlation coefficients (R and p): 
log N$_{H}$ (R=-0.30, p=0.19), 
log L$_{X}^{obs}$ (R=-0.29, p=0.21), 
log L$_{X}^{int}$ (R=-0.20, p=0.40), 
log $\lambda_{obs}$ (R=-0.29, p=0.22),  
and log $\lambda_{int}$ (R=-0.21, p=0.37).

In \citet{Riffel+22} we have found that there is a positive correlation between the bolometric luminosity of the AGN with \maM, however, if we consider only the LLAMA sample, no correlation is found (R=-0.44, p=0.0542), but when including the lower luminosity Gemini NIFS survey sources \citep[AGNIFS survey][]{Riffel+17,Riffel+22} fitted with the XSL models as described in \S~\ref{popstar}, a positive correlation is observed (for LLAMA+AGNIFS we found R=0.37, p=0.0234). Thus, our results reinforce the conclusion that there is a delay between the formation of new stars and the triggering/feeding of the AGN.

\subsection{Metalicities}

We find that the mean stellar metallicity of the AGN hosts is lower than those of the controls. This result is in agreement with the finding of \citet{Riffel+23} who found that for regions R~$<$~0.5\,R$_e$ the AGN hosts metallicity has lower values when compared with their control galaxies and larger radii values. 

These results are additionally in agreement with those derived using the nebular emission.  \citet{doNascimento+22} derived the oxygen abundances in Seyfert galaxies from MaNGA, and found that the inner regions of these galaxies display lower abundances than their outer regions.  \citet{Armah+23}, for instance, studying the BASS sample \citep{Koss+22,Oh+22} have shown that the more luminous Seyfert galaxies have lower gas metallicities. 
Our stellar metallicity estimates are qualitatively in agreement with those obtained through the gas phase abundances\footnote{In the sense that the stellar metallicities are lower in AGN than in controls and that in the gas phase oxygen abundances are lower in the AGN dominated regions when compared with the SF dominated regions.}. However, it is not a direct comparison as the gas metallicity represents the recent chemical evolution of galaxies, since it depends on the ionizing source (and on the production and release of oxygen to the galaxy ISM), whereas the mean stellar metallicity is an average over the entire lifespan of galaxies, including all generations of stars. Thus, stellar and nebular metallicities indeed indicate distinct evolutionary phases and respond differently to various processes that govern the chemical evolution of galaxies \citep{Asari+07}. It is worth mentioning that the stellar and gas metallicities are derived using radically different methods that they should not be compared in quantitative terms \citep{CidFernandes+07}. There is a huge scatter on such a comparison, but the metallicities estimated using the stellar population fitting tend to be higher when compared with the nebular one \citep[see Fig.5(e) of][]{Asari+07}.

\subsection{AGN feedback energy considerations}

The AGN's accretion energy release can be approximated using its hard X-ray luminosities to infer its bolometric luminosity. Here the bolometric luminosity was calculated using the intrinsic X-ray luminosity together with the equation~\ref{lbol}.

During the typical accretion phase duration when the power emanating from the central engine remains relatively constant ($t_{\rm AGN}$), a portion of the emitted energy will interact with the molecular gas. However, this interaction  (termed radiation coupling: $\epsilon _r$) is very uncertain \citep[e.g.][]{Harrison+18}, but theoretical models of AGN feedback propose a value of $\sim$5 percent \citep[e.g.][]{DiMatteo+05}. From the above, we can roughly estimate the accretion power that could influence the molecular gas as $E_{Rad} \approx \epsilon _r\, L_{\rm bol}\, t_{\rm AGN}$, \citep[see][for details]{Rosario+18} and thus affect the star-formation in the AGN hosts. Here we adopted $\epsilon _r$ = 0.05 \citep{DiMatteo+05} and $t_{\rm AGN}$= 1~Myr \citep{Hickox+14,Schawinski+15}.

According to \citet{Rosario+18} the coupling mechanism can take several forms, including direct radiation absorption by molecular or atomic clouds, or the influence of a mechanical agent like a thermal wind or relativistic particles within a jet.  Assuming that $E_{Rad}$ is distributed across all the molecular gas within the central region of AGN hosts, then its effect on the molecular gas by comparing it to the gravitational potential energy experienced by this gas, and so, this can give us some insight if the feedback might have enough power to affect the stability of this gas parcel. 

Since the areas we are probing are small and in the central region of the galaxies (see Tab.~\ref{tab:obs_log}) the dominant factor governing gravitational potential is the baryonic matter, encompassing both stars and gas. However, the molecular gas fraction ($f_{gas}$) in the LLAMA galaxies is small and corrections for \ion{H}{i} mass are small for disc galaxies \citep[see][for details]{Bigiel+08,Rosario+18}, we can assume that the potential energy experienced by this molecular gas is dominated by the stellar mass within this region (e.g. the values we have derived for it inside the extracted area of each spectrum) and its modulus can be approximated as:

\begin{equation}
    E_{PG} \approx G\frac{M_\star\,M_{H_2}}{\eta R} (1+f_{gas})
\end{equation}
where $R$ is the radius of a circular region that will enclose the same area as the area extracted for each spectrum\footnote{ $R=\sqrt{(slit\, width \times aperture)/ \pi}$.} and $f_{gas} = \frac{M_{H_2}}{(M_{H_2} + M_\star)}$ with $M_{H_2} = \alpha_{CO}\times$ L$'_{CO}$. We have followed \citet{Rosario+18} and adopted  $\alpha_{CO} = 1.1\, M{\odot}pc^{-2}/(K km s^{-2})$ which is the most adequate value for metal-rich nearby galaxies \citep{Sandstrom+13}. We assumed $\eta = 1$, which means that the gas is uniformly distributed in a disk of size $R$ within a uniform gravitational potential. Finally, we have corrected the L$'_{CO}$ for aperture size\footnote{L$'_{CO}$ corrected = L$'_{CO}$ (Area NIR aperture / Area Radio Aperture).} using the half-power beamwidths quoted by \citet{Rosario+18} as the radius of the CO observations. Finally, for $M_\star$ we have used the values quoted in Tab.~\ref{tab:spfits}.

\begin{figure}
\includegraphics[scale=0.5]{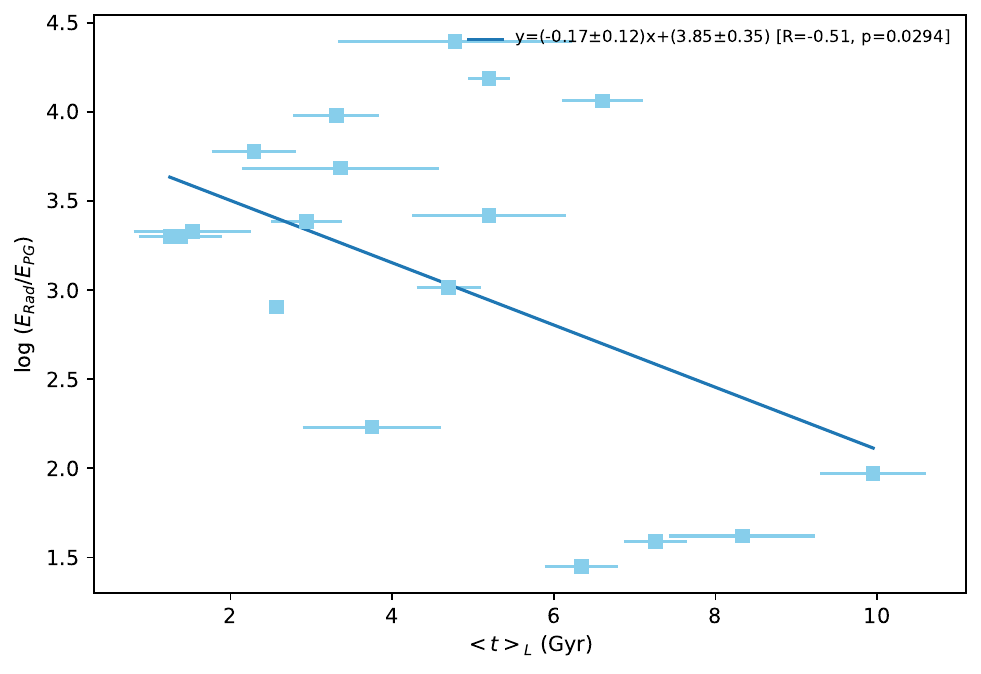}
\includegraphics[scale=0.5]{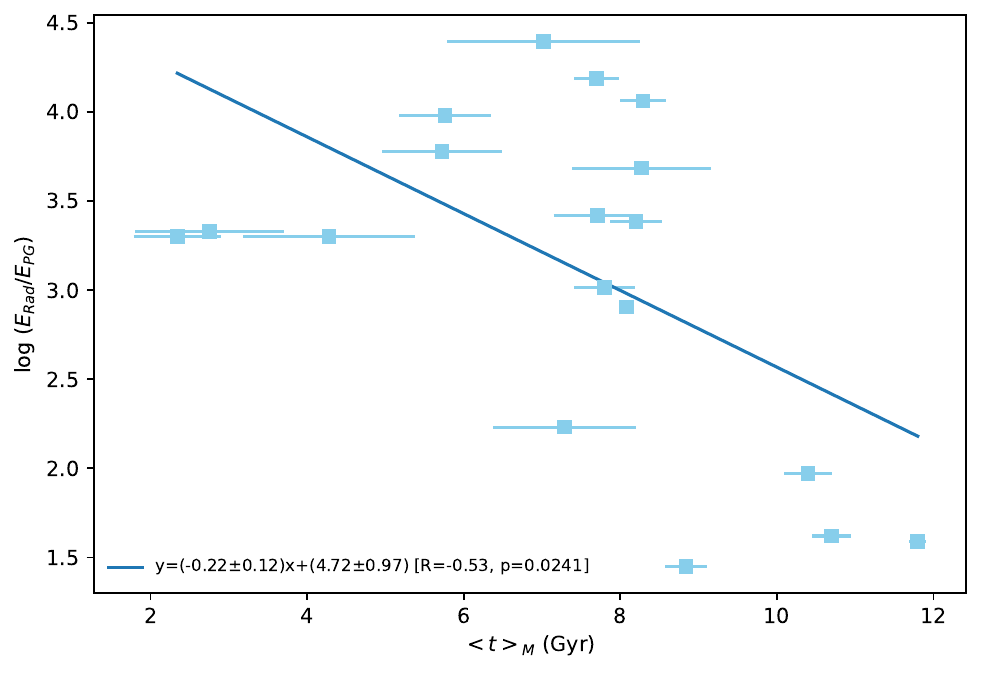}
\caption{Comparison of the AGN power with the \maL (top) and \maM(bottom).}
\label{fig:AGN_SPAges}
\end{figure}

Taking the above assumptions as our baseline, since log$(E_{Rad}$/$E_{PG}) > 0$ for all the sources (in fact, it ranges from 1.5 to 4.5 dex; Fig.~\ref{fig:AGN_SPAges}) it means that the energy injected by the AGN is enough to dynamically disturb the molecular gas within our aperture.  The  $E_{Rad}$/$E_{PG}$ ratio is compared with the AGN hosts \maL\ (top panel) and \maM\ (bottom panel) in Fig.~\ref{fig:AGN_SPAges}, where a anti-correlation is observed with the stellar population mean age ages (with Pearson coefficients R=-0.51 and p=0.0294 for \maL\ and R=-0.53 and p=0.0241 for \maM). This can be interpreted as the fact that the AGN is affecting the star formation in these galaxies, in the sense that more energetic AGN (log$(E_{Rad}$/$E_{PG}) \gtrsim 3$) tend to host younger nuclear stellar populations \maL $\lesssim$4Gyr ( \maM$\lesssim$7Gyr) or that these galaxies do have a more extended SF history since the LLAMA AGN do no have a larger amount of molecular gas when compared with their controls \citep{Rosario+18}.

It is worth mentioning that the trend observed here might be a caveat of the correction for aperture of the L$'_{CO}$, which assumes a uniform distribution of the molecular gas within the APEX beam collecting area. For instance, high-resolution ALMA observations of a volume-limited sample of X-ray selected AGN part of the Galactic Activity, Torus, and Outflow Survey (GATOS) revealed the imprint of AGN feedback in the central gas concentration of Seyfert galaxies. The GATOS AGN with higher luminosity show more outflows and more molecular gas deficits, supporting a scenario in which AGN winds are more likely to push away the gas in the center as the AGN luminosity increases \citep[][]{Garcia-Burillo+21}.The same is also observed in hot molecular gas \citep[\h2,][]{Riffel+23a}. To confirm the results obtained for the LLAMA AGN sample, we need resolved observation of the cold gas with ALMA.

\subsection{The role played by the recycled material in AGN fuelling}

Basically to trigger or sustain an AGN what is required is that enough fuel reaches the accretion disc. Such fuel (e.g. gas) can have an external origin, coming from mergers or tidal interactions \citep[e.g.][and references therein] {Raimundo+23,Araujo+23,Rembold+23}. In the case of the LLAMA AGN, \citet{Davies+17} has studied the environment of these sources. They mostly are in small or large groups, so external gas could be stripped out of other galaxies or minor mergers could play some role. However, this gas will unlikely directly feed the AGN, since it would have to reach the inner region of the galaxy. Such gas could perturb the gas already in the galaxy (LLAMA galaxies are mostly spirals, see Tab~\ref{tab:obs_log}, thus having enough recycled + pristine gas to keep SF and AGN activity \citep{Vazdekis+96}), making it reach the central region more efficiently, however, while it moves inwards it would also form stars. Finally, \citet{Davies+17} concluded that the LLAMA AGN do not show evidence of environmental dependence.

The observational results presented here show that AGN do have a larger fraction of intermediate-age stellar populations when compared with their matched analogues.  
Such an intermediate age population is dominated by massive ($\rm M\simeq 1.5 - 10\,M_{\odot}$; 200~Myr $\lesssim t \lesssim $ 2~Gyr) evolved stars \citep[e.g.][in the RGB and TP-AGB phases]{Maraston+05,Dottori+05,Salaris+14,Riffel+07,Riffel+08,Riffel+15,Riffel+22,Riffel+23,Dahmer-Hahn+22}, that do release a high amount of enriched material to the nuclear environment. 

Very recently \citet{Choi+23} have used the suite of cosmological hydrodynamical simulations of massive galaxies and supermassive black hole formation presented in \citet{Choi+17} to investigate the origins of the gas-accreted by the SMBH. They have found that recycled gas generated within the host galaxies, e.g. ejected from evolved stars (SN and AGB stars) contributes, on average, with $\sim$40\%  of the fuel for AGN. This gas also contributes to the formation of new stars but in a lower fraction ($\sim$ 20\%), being the external gas the major source of fuel for the formation of new stars. In their simulations, in-situ star formation has a higher contribution from pristine and cosmic web accretion compared to black hole feeding.

Nevertheless, they argue that this does not imply that a higher proportion of the overall recycled gas released by evolved stars contributes directly to the feeding of the black holes. In reality, given that the total gas used in star formation far exceeds the gas accreted by black holes, the sheer volume of recycled gas consumed by star formation greatly surpasses the quantity acquired by black holes. However, they show that the relatively higher proportion of recycled gas within the black hole mass budget, as compared to that attributed to star formation, suggests that recycled gas is notably efficient in feeding the SMBH, surpassing the other sources of gas entering the host galaxy. This is because this high metallicity gas cools down very efficiently and because it is abundantly released very close to the centre of the galaxy. Finally the lower (but still high, \mzL\ $\,>0.4\,Z_{\odot}$) stellar metallicity found in the AGN when compared with the control galaxies can be attributed to the fact that spiral galaxies do have a high amount of gas that has never been converted into stars \citep{Vazdekis+96}. This associated with the fact that molecular gas is very turbulent in more luminous AGN \citep{Garcia-Burillo+21}, thus making the mix of the low and high metallicity gas more efficient in AGN than in the control galaxies. Therefore diluting the overall gas phase metallicity in AGN.

The gas released by the intermediate age stars has a low velocity (a few hundred \kms) and is accreted together with the gas already available in the central region \citep[e.g.][]{Cuadra+06,Davies+07,Riffel+22,Riffel+23}. It has been demonstrated in previous works that this extra amount of gas will trigger the AGN or make it more powerful and/or form new stars \citep[e.g.][and references therein] {Ciotti+07,Ciotti+10,Leitner+11,Segers+16,Salvador-Rusinol+20,Salvador-Rusinol+21,Benedetti+23,Riffel+23}.

The larger fractions of stellar material released by the AGN when compared with the controls agree with the results found in the simulations, in the sense that besides being more abundant on AGN than in the controls, this gas is released very close to the AGN and the stars provide an efficient angular momentum sink making the gas falling more efficiently towards the SMBH \citep{Hopkins+12}. 
In this way, this recycled material acts as a "seasoning," making the supermassive black hole "food" more flavourful.

\section{Conclusions}\label{sec:conclusions}

We have used near-infrared spectroscopy to study the stellar population of the complete LLAMA sample (20 AGN) and compared them with their matched analogues (19 inactive galaxies). Therefore we have employed the new {\sc xsl} models together with the \st\ code. Our main conclusions can be summarised as follows:

\begin{enumerate}
    \item The star formation history of the sources is very complex, presenting many episodes of star formation during their lifetimes. 

     \item Optical and NIR properties derived for type~2 AGNs and controls are in very good agreement. For the NIR we have also fitted the type~1 sources and provided their stellar properties. 

    \item AGN hosts have higher fractions of intermediate-age, less metallic stellar populations when compared with the control sample. The AGN are more affected by reddening and, as expected, only they require significant fractions of FC and HD components to fit their underlying continuum.  We have obtained mean/median values of the \maL\   (4.7/4.5~Gyr for AGNs and  7.5/6.9~Gyr for the controls) as well as in the mean/median values for \maM\ (7.5/8.0~Gyr for AGNs and  10.2/10.2~Gyr for the controls).

   \item  No correlations between host galaxy gas properties (e.g. CO luminosity, hot \h2\ luminosity, \oiii\ luminosity) have been observed with age. 

 \item The  $E_{Rad}$/$E_{PG}$ ratio for AGN is compared with the \maL\ and an anti-correlation is observed with the stellar population mean age (with Pearson coefficients R=-0.48 and p=0.043). This suggests that the AGN is affecting the star formation in these galaxies, in the sense that more energetic AGNs (log$(E_{Rad}$/$E_{PG}) \gtrsim 3$) tend to host younger nuclear stellar populations \maL $\lesssim$4Gyr ( \maM$\lesssim$7Gyr).

   \item The role played by the stellar recycled material in AGN fuelling is discussed.  We found that the recent ($t<$2~Gyr) returned (recycled) stellar mass is higher in AGN than in the controls. We discuss and provide evidence to models that support that AGN feeding is, at least, partially driven by the recycled material originating from stellar evolution. 
    
\end{enumerate}

In general, our results point towards the fact that the central region of AGN are dominated by young to intermediate-age stellar populations when compared with their control galaxies. We also provide evidence that the mass loss of stars would be enough to feed the AGNs, thus providing observational constraints for models that predict that AGN feeding is primarily due to the recycled gas from
dying stars.



\section*{Acknowledgments}
We thank the anonymous referee for useful suggestions that helped improve the paper. 
We thank Ena Choi, Roberto Cid Fernandes, and Natalia Vale Asari for the useful discussion.
RR acknowledges support from the Fundaci\'on Jes\'us Serra and the Instituto de Astrof{\'{i}}sica de Canarias under the Visiting Researcher Programme 2023-2025 agreed between both institutions. RR, also acknowledges support from the ACIISI, Consejer{\'{i}}a de Econom{\'{i}}a, Conocimiento y Empleo del Gobierno de Canarias and the European Regional Development Fund (ERDF) under grant with reference ProID2021010079, and the support through the RAVET project by the grant PID2019-107427GB-C32 from the Spanish Ministry of Science, Innovation and Universities MCIU. This work has also been supported through the IAC project TRACES, which is partially supported through the state budget and the regional budget of the Consejer{\'{i}}a de Econom{\'{i}}a, Industria, Comercio y Conocimiento of the Canary Islands Autonomous Community. RR also thanks to Conselho Nacional de Desenvolvimento Cient\'{i}fico e Tecnol\'ogico  ( CNPq, Proj. 311223/2020-6,  304927/2017-1 and  400352/2016-8), Funda\c{c}\~ao de amparo \`{a} pesquisa do Rio Grande do Sul (FAPERGS, Proj. 16/2551-0000251-7 and 19/1750-2), Coordena\c{c}\~ao de Aperfei\c{c}oamento de Pessoal de N\'{i}vel Superior (CAPES, Proj. 0001). 

LGDH acknowledges support by National Key R\&D Program of China No.2022YFF0503402.
L.M. thanks FAPESP (grant 2022/03703-1) and CNPQ (grant 306359/2018-9) for partial funding of this research.
R.AR. thanks financial support from FAPERGS (21/2551-0002018-0), CNPq (303450/2022-3, 403398/2023-1, 441722/2023-7) and CAPES ( 88887.894973/2023-00). A.R.A acknowledges Conselho Nacional de Desenvolvimento Cient\'{\i}fico e Tecnol\'ogico (CNPq) for partial support to this work through grant 312036/2019-1. MT thanks the support from CNPq (process 312541/2021-0). CRA and AA acknowledge funding from the State Research Agency (AEI-MCINN) of the Spanish Ministry of Science and Innovation under the grant ``Tracking active galactic nuclei feedback from parsec to kiloparsec scales'', with reference PID2022-141105NB-I00.
\section*{Data Availability}

The data will be made available under reasonable request. 

\bibliographystyle{mnras}
\bibliography{ms_riffel_Rev1}




\includepdf[pages=-]{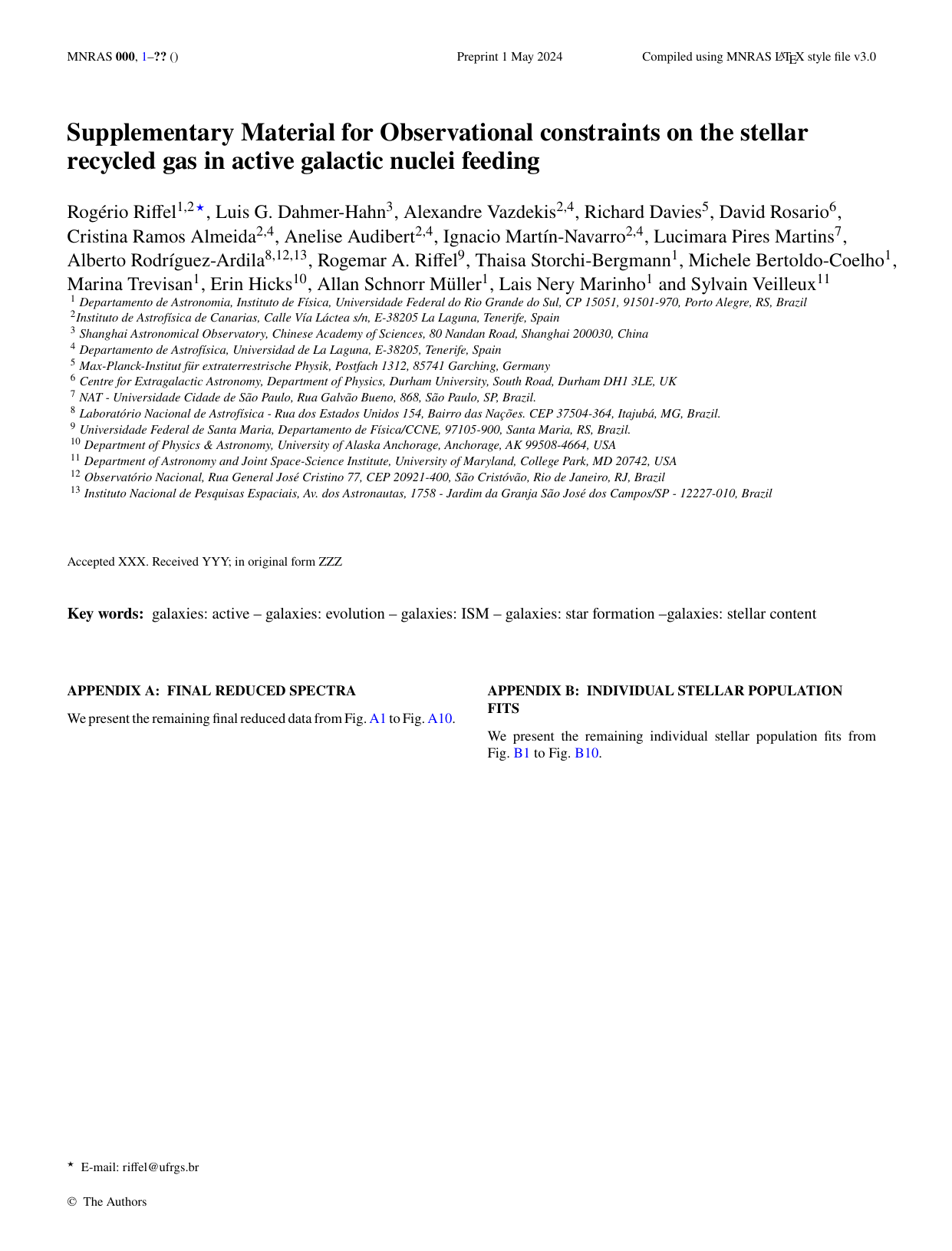}

\end{document}